\documentclass[preprint]{elsarticle}

\usepackage[utf8]{inputenc}
\usepackage{todonotes}
\usepackage{url}
\usepackage{amsmath}
\usepackage{amssymb}
\usepackage{xspace}
\usepackage[title]{appendix}
\usepackage[capitalize]{cleveref}
\usepackage{siunitx}
\usepackage{pdflscape}
\usepackage{chemformula}
\usepackage{multirow}
\usepackage{booktabs}
\usepackage{subcaption}
\usepackage[section]{placeins}
\usepackage{lineno}

\crefname{appsec}{Appendix}{Appendices}

\numberwithin{equation}{section}

\graphicspath{{./img/}}

\newcommand{\dune}{\textsc{Dune}\xspace}
\newcommand{\dumux}{DuMu\textsuperscript{x}\xspace}

\renewcommand{\div}{\nabla\cdot\!}
\newcommand{\divwrt}[1]{\nabla_{#1}\cdot\!}
\newcommand{\grad}{\nabla\!}

\newcommand\overequal[1]{\mathrel{\overset{\makebox[0pt]{\mbox{\normalfont\tiny\sffamily #1}}}{=}}}

\renewcommand{\vec}[1]{\boldsymbol{#1}}

\begin{document}

\title{A new and consistent well model for one-phase flow in anisotropic porous media using a distributed source model}
\author[1]{Timo Koch\corref{cor1}}
\ead{timo.koch@iws.uni-stuttgart.de}
\author[1]{Rainer Helmig}
\author[1]{Martin Schneider}

\cortext[cor1]{Corresponding author}
\address[1]{Department of Hydromechanics and Modelling of Hydrosystems, University of Stuttgart, Pfaffenwaldring 61, 70569 Stuttgart, Germany}

\begin{abstract}
A new well model for one-phase flow in anisotropic porous media is introduced,
where the mass exchange between well and a porous medium is
modeled by spatially distributed source terms over a small neighborhood region. To this end,
we first present a compact derivation of the exact analytical solution for an arbitrarily
oriented, infinite well cylinder in an infinite porous medium with anisotropic permeability tensor in $\mathbb{R}^3$, for constant
well pressure and a given injection rate, using a conformal map. The analytical solution motivates the choice of a
kernel function to distribute the sources. The presented model is independent
from the discretization method and the choice of computational grids.
In numerical experiments, the new well model is shown to be consistent and robust
with respect to rotation of the well axis, rotation of the permeability tensor, and different anisotropy ratios. Finally, a comparison with
a Peaceman-type well model suggests that the new scheme leads to an increased accuracy
for injection (and production) rates for arbitrarily-oriented pressure-controlled wells.
\end{abstract}


\maketitle

\section{Introduction}
Well modeling is essential for various engineering applications, as for example reservoir simulation, geothermal energy production or energy storage, where injection or extraction processes strongly influence the flow behavior. Usually the well geometry is not explicitly resolved in the mesh but instead modeled as a line source with given extraction or injection rate. However, this simplified approach introduces singularities, meaning that the logarithmic solution profiles are undefined at the center-line of the well.
This leads to a significant deviation between numerical and analytical solution in the near-well region.
For a better approximation, locally refined meshes around the wells are needed, which however deteriorate efficiency and is therefore often not suitable for field-scale simulations, especially when multiple wells are present. Similar issues are encountered for the modeling of vascularized biological tissue perfusion~\citep{Koch2018b,cattaneo2014computational} or the modeling of root water update~\citep{Koch2018a,Doussan1998}.

A common approach is the use of well-index-based well models. Such well models aim to find a relation between well rate, bottom hole pressure and numerically calculated pressure (well-block pressure) for each cell (grid-block) that contains the well. In reservoir engineering such a relation is denoted as well index. The first theoretical derivation of a well index for two-dimensional structured uniform grids with isotropic permeability has been presented by Peaceman~\citep{Peaceman1978}. He has shown that the well-block pressure differs from the areal averaged analytical pressure and introduced a new relation by using an equivalent well radius. The equivalent well radius is defined as the distance (relative to the well location) at which the analytical and numerical pressures are equal. A generalization for structured two-dimensional non-square grids ($\Delta x \not= \Delta y$) and anisotropic but diagonal permeability tensors has been presented by the same author in~\citep{Peaceman1983}. However, these well models are restricted to uniform grids where the well is oriented with one of the grid axes. Furthermore, since the effective well radius relates numerical and analytical pressure values, the well index has to be calculated depending on the used discretization scheme. Thus, Peaceman's well model is only valid for a cell-centered finite difference scheme with 5-point stencil. Discussion of other discretization schemes can be found in~\citep{chen2009well}, again with the restriction of two-dimensional grids. Enhancements include, among others, three-dimensional slanted wells~\citep{Alvestad1994,Aavatsmark2003}, Green's functions for the computation of well indices~\citep{Babu1989,Babu1991,Wolfsteiner2003}, or the singularity subtraction method to obtain smooth solutions in the near-well region~\citep{hales1997improved,Gjerde2018}.

In this work, a new approach for obtaining more accurate source term for a given well bottom hole pressure is presented. The new model is,
in contrast to most of the existence methods, independent of the discretization scheme and can be used for general unstructured grids.
Additionally, the presented method is not restricted to diagonal tensors and thus works for general anisotropic permeabilities.
In~\cref{sec:isotropic}, we derive a well model, initially for isotropic porous media, for which the fluid mass injected by a well is distributed
over a small neighborhood around the well, using kernel functions. The derivation follows the idea recently presented in~\citep{Koch2019a}, however
we herein discuss the case without membrane or casing. The model yields a pressure solution without singularity, from which the source term
can be reconstructed using a relation found with the analytical solution for the case of an infinite well in an infinite medium. The model generalizes to
more complex problems due to the superposition principal valid for the Laplace operator, a linear operator~\citep{Koch2019a}.
In~\cref{sec:aniso}, the model is generalized to porous media with general anisotropic permeabilities, based on an analytical solution constructed
in~\cref{sec:anasol} using a series of coordinate transformations. We shown that the general model reduces to the
model derived in~\cref{sec:isotropic} for isotropic permeabilities.
Finally, the new well model is analyzed with several numerical experiments in~\cref{sec:numeric}. The results indicate
that the model is consistent for different anisotropy ratios, robust with respect to rotations of the well relative to the computational grid,
and to rotations of an anisotropic permeability tensor. A comparison with a Peaceman-type well model in a setup with
a K-orthogonal grid and an embedded slanted well suggests that the new model more accurately approximates the fluid exchange between well and rock matrix.

\section{A well model with distributed source for isotropic media}
\label{sec:isotropic}

First, we derive a well model with distributed source for porous media with isotropic permeability tensor,
not including a well casing.
The derivation follows~\citep{Koch2019a}, where a casing is included in form of a membrane in a different but related application,
that is modeling fluid exchange between the vascular system and the embedding biological tissue.
Stationary single-phase flow around a well with radius $r_\omega$, in an isotropic porous medium with permeability $k$,
can be described by the following flow equation
\begin{equation}
\label{eq:flow}
  - \div \left( \frac{\rho}{\mu} k \grad p \right) = q \Phi_\Lambda \quad \text{in } \Omega,
\end{equation}
where $p$ is the fluid pressure, $\rho$ the fluid density and $\mu$ its dynamic viscosity.
Denoted by $\Phi_\Lambda$ is a set of kernel functions $\Phi_{\Lambda_i}$
that distribute $q$ (\si{\kg\per\s\per\m}) over a small tubular support region, $\mathcal{S}(\Phi_{\Lambda_i})$,
with radius $\varrho(s)$, around a well segment $i$, such that $\Phi_{\Lambda_i} = 0$ outside the support region.
We choose kernel functions $\Phi_{\Lambda_i}(s)$ with the property
\begin{equation}
\label{eq:integral_kernel}
\int\displaylimits_0^{L_i}\!\int\displaylimits_0^{2\pi}\!\int\displaylimits_0^{\varrho(s)} \Phi_{\Lambda_i} r\;\text{d}r\text{d}\theta\text{d}s = L_i,
\end{equation}
where $r$, $\theta$, $s$ are the radial, angular, and axial coordinate in a segment-local cylinder coordinate system, and $L_i$ is the length of segment $i$.

Assume a radially symmetric zone (distance $\delta > r_\omega$ from center-line) around the well, with $ p_{\delta}$ denoting
the pressure at distance $\delta$ from the well center-line, and constant fluid density and viscosity.
Then, the pressure for $r_\omega < r < \delta$ is described by the analytical solution
\begin{equation}
p(r) = - \frac{\mu}{k\rho}\frac{q}{2\pi}\ln r + C.
\end{equation}
The constant $C$ is determined by fixing a well pressure, $p_\omega$,
\begin{equation*}
p_\omega = p(r_\omega) = -\frac{\mu}{k\rho}\frac{q}{2\pi}\ln r_\omega + C \quad \Rightarrow C = p_\omega + \frac{\mu}{k\rho}\frac{q}{2\pi}\ln r_\omega,
\end{equation*}
so that
\begin{equation}
p(r) = - \frac{\mu}{k\rho}\frac{q}{2\pi}\ln \left( \frac{r}{r_\omega} \right) + p_\omega.
\end{equation}
Consequently, the source term can be expressed in terms of $p_\omega$ and $p_{\delta}$ as
\begin{equation}
\label{eq:source}
q = 2\pi r_\omega\frac{\rho k}{\mu} \frac{(p_\omega - p_{\delta})}{r_\omega \ln \left( \frac{\delta}{r_\omega} \right)}
\end{equation}
We choose a simple kernel function which regularizes the pressure solution for $r \leq \varrho$,
\begin{equation}
  \Phi^\text{const}(r) = \begin{cases}
    \frac{1}{\pi\varrho^2} & r \leq \varrho,\\
    0 & r > \varrho, \end{cases}
\end{equation}
where $\varrho \leq \delta$.
The pressure for $r < \rho$ can be obtained by integration from \cref{eq:flow}, yielding
\begin{equation}
p(r) = \begin{cases}
  -\frac{\mu}{k\rho}\frac{q}{2\pi} \left[ \frac{r^2}{2\varrho^2} + \ln\left(\frac{\varrho}{r_\omega}\right) - \frac{1}{2} \right] + p_{\omega} & r \leq \varrho,\\
  -\frac{\mu}{k\rho}\frac{q}{2\pi} \ln\left(\frac{r}{r_\omega}\right) + p_{\omega}& r > \varrho.
   \end{cases}
\label{eq:ana_constkernel}
\end{equation}
\begin{figure}
\centering
\includegraphics[width=0.9\textwidth]{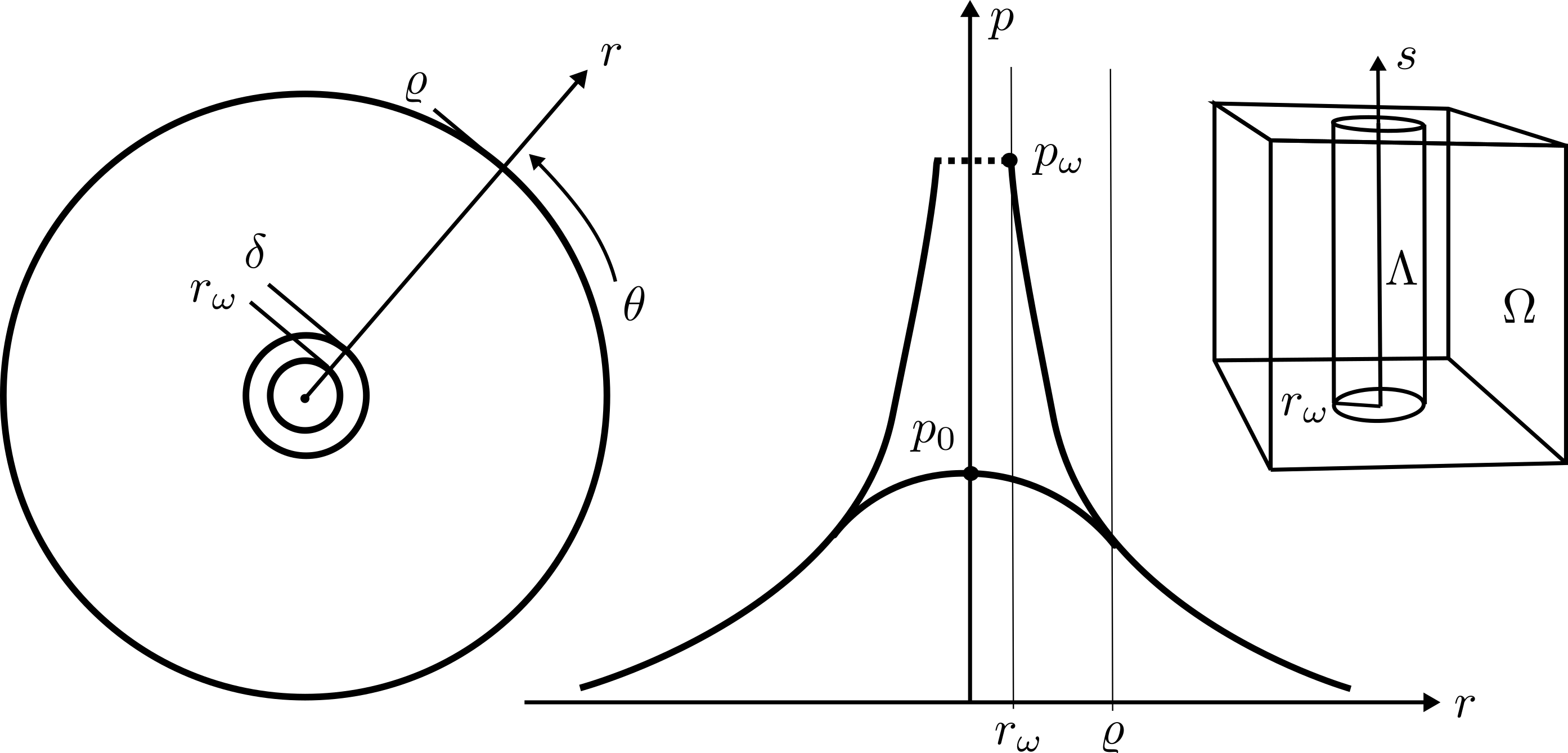}
\caption{An infiltration scenario for an isotropic porous medium.
Schematic representation of the introduced symbols. An infinite well with radius $r_\omega$,
center-line $\Lambda$ with local cylindrical coordinate system $(r, \theta, s)$ is embedded in the
porous domain $\Omega$. The kernel function with radius $\varrho$ regularizes the pressure solution which can then be
evaluated at $r=0$: $p(r=0) = p_0$.}
\label{fig:isosolution}
\end{figure}
\Cref{fig:isosolution} graphically explains the most important symbols introduced in this section.
As the regularized pressure can be evaluated at the well center-line we can reformulate~\cref{eq:source},
\begin{equation}
\label{eq:sourcewithxi}
q = 2\pi \frac{\rho k}{\mu}(p_\omega - p_0)\Xi, \quad\text{with}\; \Xi = \frac{(p_\omega - p_{\delta})}{(p_\omega - p_0)}\frac{1}{\ln \left( \frac{\delta}{r_\omega} \right)}
\end{equation}
where $\Xi$ is the so called flux scaling factor.
The flux scaling factor can be also expressed independent of the pressure. To this end,
\cref{eq:ana_constkernel} is evaluated at $r=0$, so that $p_{\omega}$ is expressed in terms of $p_0$,
\begin{equation}
p_0 = - (p_\omega - p_0)\Xi\left[ \ln\left(\frac{\varrho}{r_\omega}\right) - \frac{1}{2} \right] + p_{\omega},
\label{eq:pt0andptw}
\end{equation}
where $q$ was replaced by inserting \cref{eq:sourcewithxi}.
It directly follows from \cref{eq:pt0andptw} that
\begin{equation}
\Xi = \left[\ln\left(\frac{\varrho}{r_\omega}\right)-\frac{1}{2} \right]^{-1}.
\label{eq:xi_const}
\end{equation}

\section{A well model with distributed source for anisotropic media}
\label{sec:aniso}

In the following section, the developed well model is extended for porous media with anisotropic permeability.
In~\cref{sec:anasol}, we derive an analytical solution for one-phase flow around an infinitely long cylindrical well embedded
in an infinite porous domain in $\mathbb{R}^3$. This derivation motivates the choice of a suitable kernel function for
anisotropic problems, presented in~\cref{sec:anisokernel}.

\subsection{Analytical solution for anisotropic permeability and slanted well}
\label{sec:anasol}

In the following section, we derive an analytical solution for one-phase flow around an infinite cylindrical well $\Gamma$
with radius $r_\omega$ in an infinite porous domain $\hat{\Omega} = \mathbb{R}^3\setminus\Gamma$ with anisotropic, homogeneous permeability.
We assume, without loss of generality, that the well axis passes through the origin of the Cartesian coordinate system,
and denote by $\vec{\psi}$ a unit vector parallel to the well signifying the well orientation. We seek an analytical expression
for the hydraulic pressure $p$ such that
\begin{equation}
\label{eq:flow_ansio}
  - \div \left( \frac{\rho}{\mu} K \grad p \right) = 0 \quad \text{in } \hat{\Omega},
\end{equation}
for a constant well pressure $p_\omega$ in \si{\Pa}
and some specific pumping rate $q$ in \si{\kilogram\per\s\per\m} given on $\partial\Gamma$. The total mass flow over the
boundary of a well segment of length $L$ is thus given by $Q = qL$.

From thermodynamic constraints, $K$ is a positive definite and symmetric, second-order tensor field.
Hence, $K$ can be decomposed such that
\begin{equation}
K = Q D Q^T,
\end{equation}
where $D = \operatorname{diag}(\lambda_1, \lambda_2, \lambda_3)$ is a diagonal matrix composed of the eigenvalues $\lambda_i$ of $K$,
$Q = [ \vec{\nu}_{K,1} \vert \vec{\nu}_{K,2} \vert \vec{\nu}_{K,3} ]$ is a rotation matrix with the corresponding eigenvectors as columns,
and $A^T$ denotes the transpose of a matrix $A$.
Further useful properties derived from the decomposition are $\operatorname{det}(K) = \lambda_1\lambda_2\lambda_3$,
where $\operatorname{det}(A)$ denotes the determinant of $A$, and $K^n = Q \Lambda^n Q^T$,
where $D^r = \operatorname{diag}({\lambda_1^r, \lambda_2^r, \lambda_3^r})$, $r \in \mathbb{R}$.

It is well known that the anisotropic one-phase flow problem can be transformed to an isotropic problem
using a coordinate transformation~\citep{Aavatsmark2003,Fitts2006,Aavatsmark2016,Peaceman1983,Bear1965}
\begin{equation}
U: \mathbb{R}^3\rightarrow \mathbb{R}^3, \vec{x} \mapsto \vec{u} = \tilde{S}\vec{x},
\end{equation}
with the stretching matrix $\tilde{S} = k_I^{1/2} K^{-1/2}$, where $k_I$ is an arbitrary scalar constant, that we
choose as $k_I = \operatorname{det}(K)^{-1/3}$ (cf. \citep{Aavatsmark2003}), rendering the transformation isochoric.
The transformation $\vec{u} = \tilde{S}\vec{x}$ deforms the well cylinder such that a cross-section
orthogonal to the transformed well direction is elliptical.
The solution to the isotropic problem
\begin{equation}
\label{eq:flow_iso}
  - \divwrt{u} \left( \frac{\rho}{\mu} k_I \grad_u p \right) = 0 \quad \text{in } \hat{\Omega}_u = U(\hat{\Omega}),
\end{equation}
is identical on two parallel planes perpendicular to the transformed (normalized) well direction,
$\vec{\psi}' = \tilde{S}\vec{\psi} \vert\vert \tilde{S}\vec{\psi} \vert\vert^{-1}$.
This motivates the rotation of the coordinate system such that the first and second axis are aligned with
the major and minor axis of the well-bore ellipse and third axis is aligned with $\vec{\psi}'$.
To determine the corresponding rotation matrix $\tilde{R}$, we need to characterize this well-bore ellipse.
The well cylinder in $x$-coordinates is given by
\begin{equation}
\vec{x}^T \Psi \vec{x} = r_\omega^2, \quad \Psi = I - \vec{\psi}\vec{\psi}^T.
\end{equation}
After stretching, the coordinate system can be rotated with the rotation matrix $R$ so that the third axis is aligned with the well direction.
Then, projecting into the plane perpendicular to the well direction yields the well-bore ellipse equation
\begin{equation}
\label{eq:wellboreellipse}
\hat{\vec{v}}^T E \hat{\vec{v}} = \hat{\vec{v}}^T P^T R \tilde{S}^{-1} \Psi \tilde{S}^{-1} R^T P \hat{\vec{v}} = r_\omega^2, \quad \tilde{S}^{-1}R^T P\hat{\vec{v}} = \vec{x}
\end{equation}
in $\hat{v}$-coordinates, where
\begin{equation}
\quad R = 2\frac{(\vec{e}_3 + \vec{\psi}')(\vec{e}_3 + \vec{\psi}')^T}{(\vec{e}_3 + \vec{\psi}')^T(\vec{e}_3 + \vec{\psi}')} - I, \quad \vec{e}_3 = \left[\begin{array}{c}0\\0\\1\end{array}\right], \quad P = \left[\begin{array}{cc}1&0\\0&1\\0&0\end{array}\right].
\end{equation}
The rotation matrix $R$ can be derived using Rodrigues' rotation formula as shown in~\cref{sec:appendixA}.
The length of the major and minor ellipse axis are found as $a = r_w \gamma_1^{-1/2}$ and $b = r_w \gamma_2^{-1/2}$,
where $\gamma_i$ are the eigenvalues of $E$, and the axis orientations are given by $\vec{\nu}_1 = P\hat{\vec{\nu}}_{E,1}$, $\vec{\nu}_2 = P\hat{\vec{\nu}}_{E,2}$,
where $\hat{\vec{\nu}}_{E,i}$ denote the corresponding eigenvectors of $E$. We assume that the eigenvalues and eigenvectors are sorted such
that $a\geq b$, and oriented such that $\vec{\psi'} = \vec{\nu}_1 \times \vec{\nu}_2$. Finally the desired rotation is given by
\begin{equation}
V: \mathbb{R}^3\rightarrow \mathbb{R}^3, \vec{u} \mapsto \vec{v} = \tilde{R}\vec{u} = \hat{R}^T R^T\vec{u},
\end{equation}
where
\begin{equation}
\hat{R} = \left[ \vec{\nu}_1 \bigg\vert \vec{\nu}_2 \bigg\vert \vec{\psi'} \right]
\end{equation}
is rotating about the well direction axis such that the coordinate system is aligned with the principal ellipse axes.

\begin{figure}
 \centering
 \includegraphics[width=1.0\textwidth]{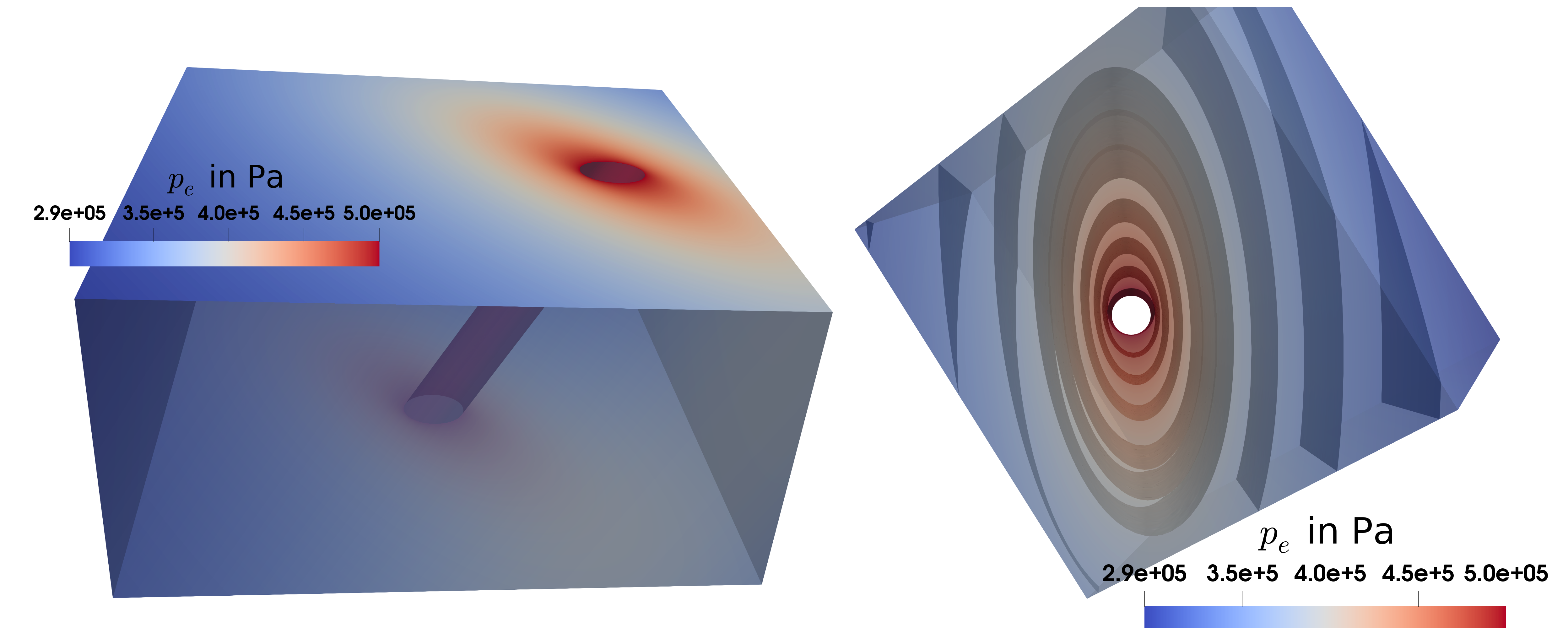}
 \caption{The analytical pressure solution for a slanted well, $r_\omega = \SI{0.1}{\meter}$, with well pressure $p_\omega = \SI{5.0e5}{\Pa}$,
          total mass injection rate $Q_\omega = \SI{115.47}{\kg\per\s}$, and anisotropic permeability tensor $K_A$.
          The top view is oriented in well direction and shows pressure contour surfaces highlighting their the elliptical shape.}
 \label{fig:analytical3d}
\end{figure}

Following the derivations from above, we now have to solve a two-dimensional isotropic Laplace problem
with boundary conditions prescribed on an ellipse. To this end,
we note that the transformation of a harmonic function $f$
(a function satisfying Laplace's equation $\Delta f = \div\grad f = 0$)
with a conformal (angle-preserving) mapping yields another harmonic function~\citep{Nehari1975} (see~\cref{sec:appendixB}).
Using a Joukowsky transformation, a conformal mapping well-known from aerodynamics~\citep{joukowsky1910},
the isotropic problem with a well with elliptic cross-sections,
can be transformed to an isotropic problem with circular cross-sections~\citep{Fitts2006}.
Transforming into the complex plane (parametrizing the well-bore ellipse plane)
\begin{equation}
Z: \mathbb{R}^3\rightarrow\mathbb{C}, \vec{v} \mapsto z = \tilde{Z}\vec{v} = [1, i, 0]\vec{v} = v_1 + iv_2,
\end{equation}
the (inverse) Joukowsky transformation
\begin{equation}
\label{eq:invjou}
T: \mathbb{C}\rightarrow\mathbb{C}, z \mapsto w = z + \sqrt{z-f}\sqrt{z+f}, \quad f = \sqrt{a^2 - b^2}
\end{equation}
transforms elliptic isobars into circular isobars, where $a$ and $b$,
$a\geq b$, are the major and minor
axis of the well-bore ellipse, as derived above. In particular, the well-bore ellipse (where $p = p_{\omega}$)
is mapped onto a circle with radius $r_\circ = a + b$. Finally, in the new coordinate system we find the (now) radially symmetric
analytical solution to problem~\cref{eq:flow_ansio}
\begin{equation}
p(w) = p_\omega - \frac{\mu}{\rho k_I} \frac{\hat{q}}{2\pi} \ln \left( \frac{\vert w \vert}{r_\circ} \right)\zeta, \quad \hat{q} = q\zeta = q\frac{ab}{r_\omega^2},
\end{equation}
where the source scaling factor $\zeta$ is necessary to recover the original source $q$ on $\partial\Gamma$. This can be derived from
simple geometric considerations as shown in~\cref{sec:appendixC}.
The $w$ corresponding to some $x \in \hat{\Omega}$ in original coordinates is obtained
by using all above-mentioned transformations after each other as follows
\begin{equation}
w = T(Z(V(U(\vec{x})))) = T(\tilde{Z}\tilde{R}\tilde{S}\vec{x}).
\end{equation}
Such a solution for a slanted well ($\SI{30}{\degree}$ with respect to vertical axis) and anisotropic permeability tensor
\begin{equation}
K_A = \left[\begin{array}{ccc} 1&0&0 \\ 0&5&4 \\ 0&4&5 \end{array}\right] \SI{1e-10}{\square\m}
\end{equation}
is visualized in~\cref{fig:analytical3d}.

\subsection{Properties of the conformal mapping}
\label{sec:conformalmap}

\begin{figure}
 \centering
 \includegraphics[width=1.0\textwidth]{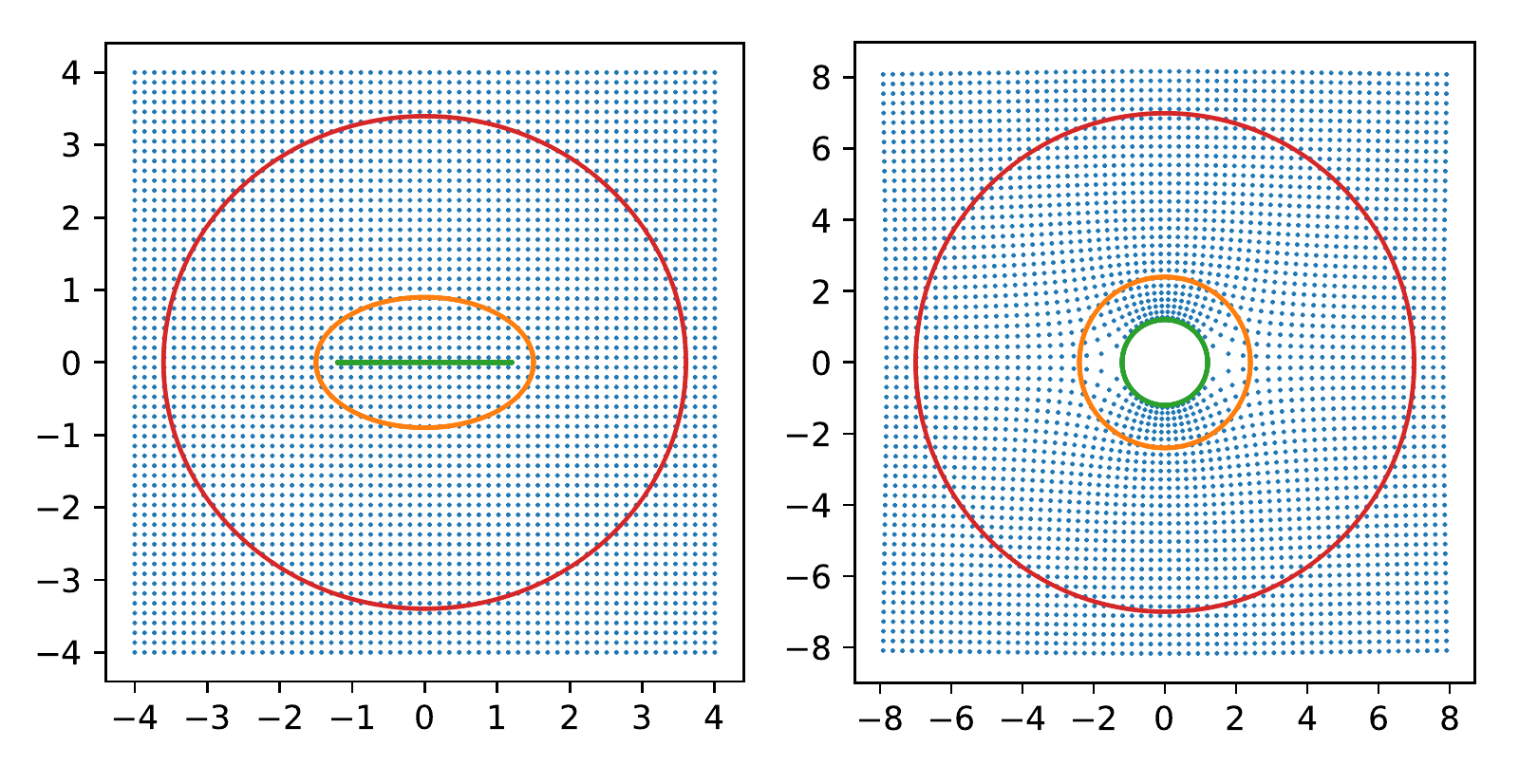}
 \caption{Visualization of the (inverse) Joukowsky transformation, $T$, exemplarily for $a=1.5$, $b=0.9$, and thus $f=1.2$.
          The points on the left are shown on the complex $z$-plane, the points on the right are shown in the complex $w$-plane, and $w = T(z)$.
          The inner circle on the right image has radius $f$. The other circles have radii of $r_\circ = a + b$ and $3r_\circ$.}
 \label{fig:joukowsky}
\end{figure}
To construct a suitable kernel function for anisotropic problems, we first have a closer look at the properties of
the employed Joukowsky transformation. The effect of the mapping $T$,~\cref{eq:invjou}, is shown in~\cref{fig:joukowsky}.
Points on the exterior of a line on the real axis between $f$ and $-f$ are mapped onto the exterior of a circle with radius $f$.
The ellipse with major axis $a$ and minor axis $b$ is mapped onto a circle with radius $r_\circ = a + b$.
Going further away from the well, the deformation due to the mapping is less and less pronounced. This matches the expectations for
the physical flow problem, since isobars at large distance from an elliptical well-bore become increasingly circular in isotropic media.

\begin{figure}
 \centering
 \includegraphics[width=0.7\textwidth]{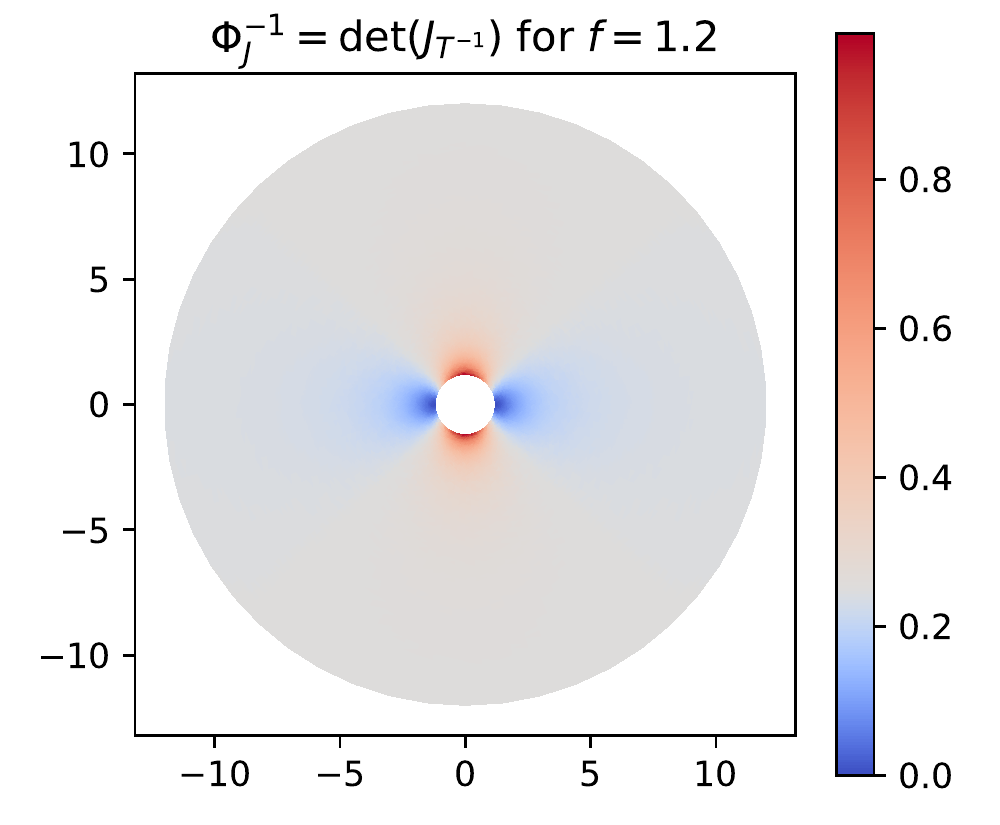}
 \caption{The determinant of the Jacobian of the transformation $z = T^{-1}(w)$ for $\vert w\vert > f$ where $f^2 = a^2 - b^2$ and $a>b$.
          Larger volume deformations only occur very locally in vicinity of the well radius $r_\circ = a + b$ and quickly converge to $0.25$
          with larger distance to the well.}
 \label{fig:detplot}
\end{figure}

The inverse transformation is given by
\begin{equation}
\label{eq:joumap}
T^{-1}: \mathbb{C}\rightarrow\mathbb{C}, w \mapsto z = \frac{1}{2}\left( w + \frac{f^2}{w} \right), \quad \vert w \vert > f,
\end{equation}
where the restriction on $\vert w \vert$ is necessary to obtain a one-to-one mapping. The transformation $T^{-1}$ can equally
be interpreted as an $\mathbb{R}^2\rightarrow \mathbb{R}^2$ mapping.
The Jacobian of the transformation $T^{-1}$ for $z = x + iy$ and $w = u + iv$ has the form
\begin{equation}
\label{eq:jou_jac}
J_{T^{-1}} = \left[\begin{array}{cc} \frac{\partial x}{\partial u} &  \frac{\partial x}{\partial u} \\
                                     \frac{\partial y}{\partial v} &  \frac{\partial y}{\partial v} \end{array}\right]
           = \left[\begin{array}{cc} \eta & \epsilon \\ -\epsilon & \eta \end{array}\right],
\end{equation}
which follows from the the Cauchy--Riemann equations~\citep{Rudin1987}. Since the transformation can be viewed
as the composition of a scaling and a rotation, it is angle-preserving. The transformation is
associated with a spatially dependent volume deformation characterized by the determinant of $J_{T^{-1}}$. Furthermore, it can be
shown that the Laplace operator behaves as follows under the transformation $z = T^{-1}(w)$,
\begin{equation}
\label{eq:laplaceoperatormapped}
  \Delta_w p = \frac{\partial^2 p}{\partial u^2} + \frac{\partial^2 p}{\partial v^2} = \left| \frac{\partial T^{-1}}{\partial w}\right|^2 \Delta_z p = \left|\operatorname{det}(J_{T^{-1}})\right| \Delta_z p
\end{equation}
by computing the derivative of the real and the imaginary part of $p$ separately, applying the chain rule and the Cauchy--Riemann equations, as
shown for completeness in~\cref{sec:appendixB}.
From~\cref{eq:laplaceoperatormapped} follows that
\begin{equation}
\label{eq:laplaceoperatormappedinv}
  \left|\operatorname{det}(J_{T^{-1}})\right|^{-1} \Delta_w p = \Delta_z p,
\end{equation}
for the transformation $w = T(z)$. The determinant can be explicitly computed,
using~\cref{eq:laplaceoperatormapped} and complex differentiation (shown in~\cref{sec:appendixD}) as
\begin{equation}
\label{eq:detfunction}
\left|\operatorname{det}(J_{T^{-1}})(w)\right| =  \left| \frac{\partial T^{-1}}{\partial w}\right|^2 = \left| \frac{\partial z}{\partial u}\right|^2 = \frac{1}{4}\left( 1 + \frac{f^4 - 2f^2\Re(w^2)}{\vert w \vert^4} \right) := \Phi_J^{-1},
\end{equation}
where $\Re(w^2)$ is real part of $w^2$, and $\vert w \vert$ the absolute value of $w$. We note that $\Phi_J$ quickly converges
to the value $4$ with increasing $\vert w \vert$, that is with increasing distance from the well. The function $\Phi_J^{-1}$ is
plotted in~\cref{fig:detplot} exemplarily for $f = 1.2$.

\subsection{A kernel function for anisotropic media}
\label{sec:anisokernel}
Instead of excluding the well domain $\Gamma$ from $\Omega = \mathbb{R}^3$ and modeling infiltration or extraction by a flux boundary condition,
we will now model the action of the well on the flow field by a spatially distributed source term, as presented for the isotropic problem,
\begin{equation}
\label{eq:flowaniso_kernel}
  - \div \left( \frac{\rho}{\mu} K \grad p \right) = q\zeta \Phi_\Lambda \quad \text{in } \Omega.
\end{equation}
From the above derivations, we know that solving~\cref{eq:flowaniso_kernel} in $w$-coordinates is straight-forward. Hence,
we choose kernel functions in $w$-coordinates and then transform to $x$-coordinates so that the pressure solution
satisfies~\cref{eq:flowaniso_kernel}. Motivated by the properties of
the Joukowsky transform (see~\cref{fig:joukowsky}),
we choose a local kernel that is constant on the annulus with inner radius $f < \varrho_i \leq r_\circ$
and outer radius $\varrho_o > r_\circ$,
\begin{equation}
\Phi_A(w) = \begin{cases}
  \frac{1}{\pi\left(\varrho_o^2 - \varrho_i^2 \right)} & \varrho_i \leq \vert w \vert \leq \varrho_o,\\
  0 & \text{elsewhere}. \end{cases}
\end{equation}
In $w$-coordinates, we can find a solution to the problem
\begin{equation}
\label{eq:flowaniso_kernel_w}
  - \Delta_w p = \hat{q}\frac{\mu}{\rho k_I} \Phi_A \quad \text{in } \Omega_w = T(Z(V(U(\Omega)))),
\end{equation}
for a given constant well pressure $p_\omega$, $\hat{q} = q\zeta$ and constant density and viscosity.

By means of integration (cf.~\citep{Koch2019a}), we get
\begin{equation}
\label{eq:ana_w}
p(w) = \begin{cases}
  p_{\omega}-\frac{\mu}{K\rho}\frac{\hat{q}}{2\pi} \left[ \frac{(\vert w\vert^2 - \varrho_o^2)}{2\xi^2} - \frac{\varrho_i^2}{\xi^2}\ln\left(\frac{\vert w\vert}{\varrho_o}\right) + \ln\left(\frac{\varrho_o}{r_\omega}\right) \right] & \varrho_i \leq \vert w\vert \leq \varrho_o,\\
  p_{\omega}-\frac{\mu}{K\rho}\frac{\hat{q}}{2\pi} \left[ -\frac{1}{2} - \frac{\varrho_i^2}{\xi^2}\ln\left(\frac{\varrho_i}{\varrho_o}\right) + \ln\left(\frac{\varrho_o}{r_\omega}\right) \right] & \vert w\vert < \varrho_i, \\
  p_{\omega}-\frac{\mu}{K\rho}\frac{\hat{q}}{2\pi} \ln\left(\frac{\vert w\vert}{r_\omega}\right) & \vert w\vert > \varrho_o,
   \end{cases}
\end{equation}
where $\xi^2 = \varrho_o^2 - \varrho_i^2$. This shows that outside the kernel support region ($\vert w\vert > \varrho_o$), we obtain the exact
analytical solution derived in~\cref{sec:anasol}. Moreover, the source term can be reformulated, cf.~\citep{Koch2019a},
\begin{equation}
\hat{q} = 2\pi \frac{\rho k_I}{\mu}(p_\omega - p_0)\Xi, \quad \Xi = \left[\ln\left(\frac{\varrho_o}{r_\omega}\right)-\frac{1}{2} - \frac{\varrho_i^2}{\xi^2}\ln\left(\frac{\varrho_i}{\varrho_o}\right) \right]^{-1},
\end{equation}
where $p_0 := p(\vert w\vert = 0) = p(\vert w\vert = \varrho_i)$ is the fluid pressure evaluated on the well center-line.
Note that for $f = 0$ and $\varrho_i = f = 0$, the isotropic solution with for a circular constant kernel (\cref{eq:ana_constkernel,eq:xi_const}) is obtained.
From the transformation of the Laplace operator,~\cref{eq:laplaceoperatormappedinv}, we see that the problem
\begin{equation}
\label{eq:flowaniso_kernel_z}
  - \Delta_z p = \hat{q}\frac{\mu}{\rho k_I} \Phi_A \Phi_J  \quad \text{in } \Omega_z = Z(V(U(\Omega))),
\end{equation}
with altered kernel function $\Phi_\Lambda = \Phi_A\Phi_J$ is equivalent to~\cref{eq:flowaniso_kernel_w}.

\begin{figure}
 \centering
 \includegraphics[width=0.8\textwidth]{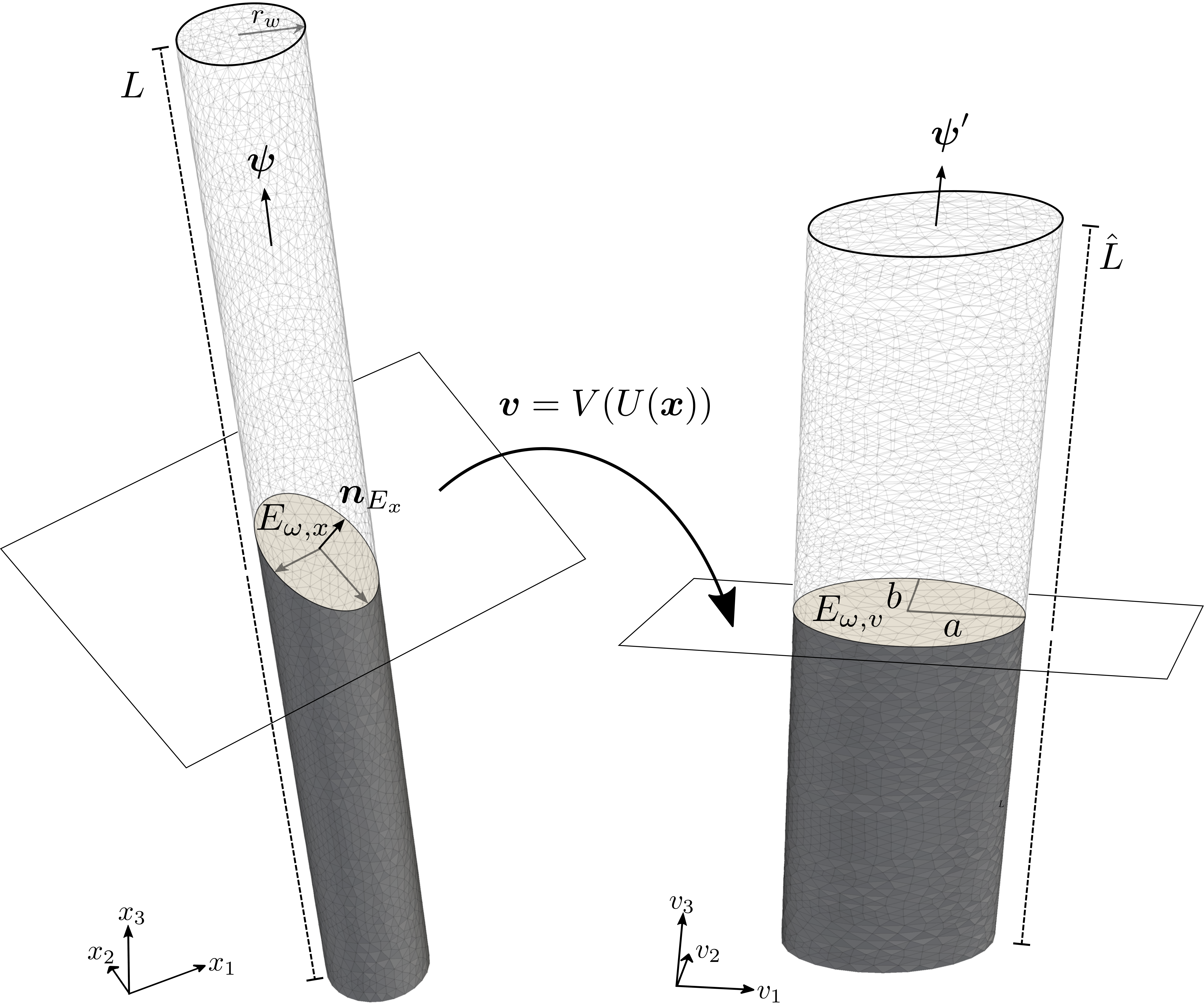}
 \caption{Visualization of the coordinate transformation $\vec{v} = V(U(\vec{x})) = \tilde{R}\tilde{S}\vec{x}$.
          The ellipse $E_v$ is orthogonal to the well direction $\vec{\psi}'$ which is equal to $\vec{e}_3 = [0, 0, 1]^T$ in $v$-coordinates.}
 \label{fig:transformation}
\end{figure}
The transformation $T^{-1}$ changes shape of the
kernel support $\mathcal{S}(\Phi_A)$ from an annulus to an ellipse $E_{\Phi,v}$. Inverting $Z$ extrudes the solution along the well center-line,
and inverting the rotation and stretch described by $V$ and $U$ results in a kernel support region in the shape of
an elliptic cylinder. Moreover, each ellipse $E_{\Phi,v}$ with normal vector $\vec{e}_3$ is transformed to an ellipse $E_{\Phi,x}(s)$, that is
the intersection the elliptic cylinder with a plane with the normal vector $\vec{n}_{E_x} = \tilde{S}\tilde{R}^T\vec{e}_3$,
centered at $s$ on the well center-line. The transformation and the normal vector $\vec{n}_{E_x}$ are visualized in~\cref{fig:transformation}.
We note that if none of the principal axes of the permeability tensor are aligned with the well direction,
$\vec{n}_{E_x}$ is not parallel to the well direction $\vec{\psi}$ in $x$-coordinates.
The integral of the right-hand side of~\cref{eq:flowaniso_kernel}
for a well segment $\Lambda_i$ with length $L_i$ is equal to the integral over the kernel support $\mathcal{S}(\Phi_{\Lambda,i})$
which has the shape of the elliptic cylinder given by
\begin{equation}
\mathbb{E} := \bigcup\limits_{0 \leq s \leq L_i} E_{\Phi,x}(s).
\end{equation}
Using $\hat{q} = q\zeta$, and exploiting that $k_I$ was chosen such that $\operatorname{det}(\tilde{R}\tilde{S}) = 1$, it can be shown that
\begin{equation}
\label{eq:kernelintegralsegment}
\int\displaylimits_{\mathbb{E}} \hat{q}\Phi_{\Lambda,i} \text{d}x
= \!\!\!\int\displaylimits_{V(U(\mathbb{E}))}\!\!\! \hat{q}\Phi_{\Lambda,i} \text{d}v
= \int\displaylimits_0^{\hat{L}_i} \!\int\displaylimits_{E_{\Phi,v}(\hat{s})}\! \hat{q}\Phi_{\Lambda} \text{d}\hat{A}\text{d}\hat{s} = \hat{q}\hat{L}_i = qL_i,
\end{equation}
where $0 \leq \hat{s} \leq \hat{L}_i$ is a local coordinate along the transformed well direction, and the last equality is proven in~\cref{sec:appendixC}.
This is the desired property of the kernel function for the anisotropic case corresponding to~\cref{eq:integral_kernel} for the isotropic case.

\section{Numerical method}
\begin{figure}
 \centering
 \includegraphics[width=0.7\textwidth]{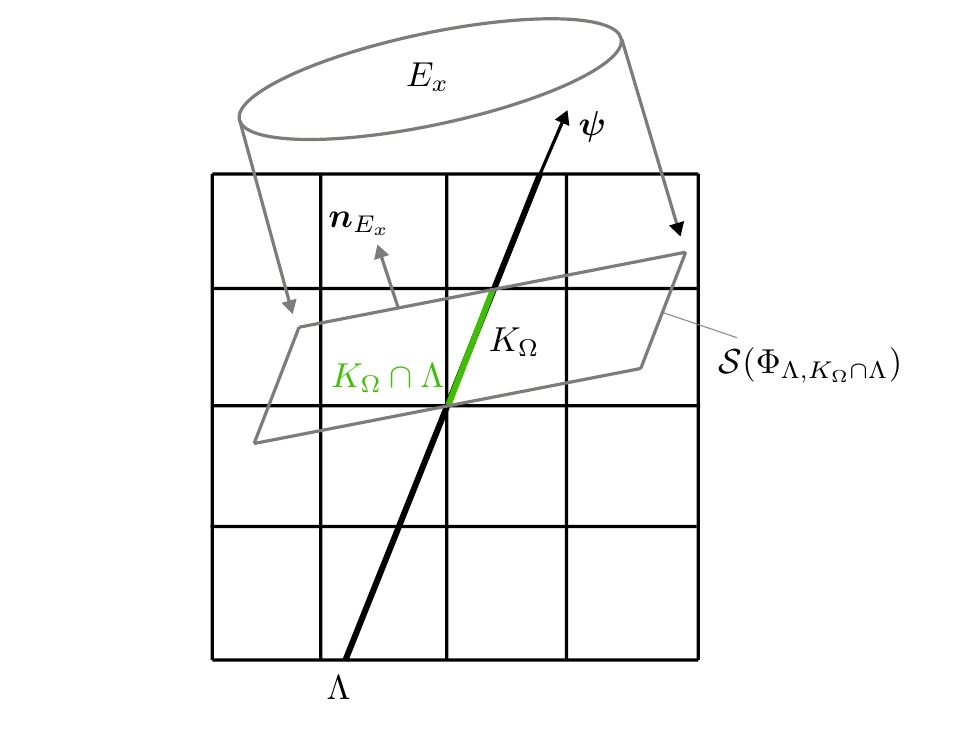}
 \caption{Visualization of the discretization process. The domain $\Omega$ is represented by a set of control volumes $K_\Omega \in \Omega_h$.
 The well center-line $\Lambda$ with direction $\vec{\psi}$ intersects with a $K_\Omega$ shown in green. The gray parallelogram is a 2D-projection of
  the elliptic cylinder that is the part of the kernel support $\mathcal{S}(\Phi_{\Lambda})$ associated with $K_\Omega \cap \Lambda$.}
 \label{fig:disc}
\end{figure}

We discretize \cref{eq:flowaniso_kernel} using a cell-centered finite volume method with multi-point flux approximation (MPFA)~\citep{Aavatsmark2002}.
The domain $\Omega$ is decomposed into control volumes $K_\Omega \in \Omega_h$
such that the computational mesh $\Omega_h$ is a discrete representation of $\Omega$.
Furthermore, each control volume boundary, $\partial K_\Omega$, can be split into a finite number of faces $\sigma\subset\partial K_\Omega$,
such that $\sigma = K_\Omega\cap L_\Omega$,
with $L_\Omega$ denoting a neighboring control volume. Integrating \cref{eq:flowaniso_kernel}
over a control volume $K_\Omega$ and applying the Gauss divergence theorem on the left hand side yields
\begin{equation}
\label{eq:disc_integral}
-\int_{\partial K_\Omega}\! \left[ \frac{\rho}{\mu} K \grad p \right]\cdot\boldsymbol{n}_{K_\Omega,\sigma}\, \text{d}A = \int_{K_\Omega}\! \hat{q} \Phi_\Lambda \,\text{d}x,
\end{equation}
where $\boldsymbol{n}_{K_\Omega,\sigma}$ is the unit outward-pointing normal on face $\sigma\subset\partial K_\Omega$.
The exact fluxes are approximated by numerical fluxes
\begin{equation}
F_{K_\Omega,\sigma} \approx -\int_{\sigma}\! \left[ \frac{\rho}{\mu} K \grad p \right]\cdot\boldsymbol{n}_{K_\Omega,\sigma},
\end{equation}
which are computed using the MPFA\nobreakdash-O method described in~\citep{Aavatsmark2002}.
The discrete source term is computed as
\begin{equation}
\label{eq:qkomegakernel}
Q_{K_\Omega} \approx \int_{K_\Omega}\! \hat{q} \Phi_\Lambda \,\text{d}x, \quad
Q_{K_\Omega} = \frac{Q_\mathcal{I}}{\vert\mathcal{I}\vert}\int_{K_\Omega \cap \mathcal{S}(\Phi_{\Lambda,\mathcal{I}})}\!\! \Phi_\Lambda \,\text{d}x,
\end{equation}
where $Q_\mathcal{I}$ is a numerical approximation of the source term integral over the intersection $\mathcal{I} = K_\Omega \cap \Lambda$,
\begin{equation}
\label{eq:discretesourceapprox}
Q_\mathcal{I} = \vert\mathcal{I}\vert 2\pi \frac{\rho k_I}{\mu}(p_\omega - p_0)\Xi,
\end{equation}
and $\mathcal{S}(\Phi_{\Lambda,\mathcal{I}})$ is the kernel support associated with $\mathcal{I}$ as depicted in~\cref{fig:disc}.
In summary, the discrete form of \cref{eq:disc_integral} is
\begin{equation}
\sum\limits_{\sigma \subset \partial K_\Omega} F_{K_\Omega,\sigma} = Q_{K_\Omega}, \quad K_\Omega \in \Omega_h.
\end{equation}
We note that due to the dependency of $Q_\mathcal{I}$ on $p_0$ the proposed method is non-local in the sense that
non-neighbor cells $M_\Omega \in \Omega_h$ (where $M_\Omega \cap K_\Omega$ is the empty set or a single point)
may have an associated degree of freedom that depends on the degree of freedom of $K_\Omega$.

\subsection{Kernel integration}

The kernel integral in \cref{eq:qkomegakernel}
\begin{equation}
\mathbb{I}_{\Phi,K_\Omega} := \int_{K_\Omega \cap \mathcal{S}(\Phi_{\Lambda,\mathcal{I}})} \Phi_\Lambda \,\text{d}x,
\end{equation}
is not easily approximated with a quadrature rule, since the intersection $K_\Omega \cap \mathcal{S}(\Phi_{\Lambda,\mathcal{I}})$, that is the intersection of
an elliptic cylinder with for example a hexahedron is difficult to compute. However, we use the same idea as in~\citep{Koch2019a},
and remark that the integral over the entire support $\mathcal{S}(\Phi_{\Lambda,\mathcal{I}})$ is known exactly; see \cref{eq:kernelintegralsegment}.
Hence, the integration problem can be reformulated as the distribution of the known integral over all intersected control volumes $K_\Omega$
weighted with the respective support volume fractions. Following~\citep{Koch2019a},
we create $n_\mathcal{I}$ integration points $\boldsymbol{x}_i \in \mathcal{S}(\Phi_{\Lambda,\mathcal{I}})$
with known volume elements $V_i$ of similar size and shape, so that
\begin{equation}
\mathbb{I}_{\Phi,K_\Omega} \approx \sum\limits_{i=1, \boldsymbol{x}_i\in K_\Omega}^{n_\mathcal{I}} V_i\Phi_\Lambda(\boldsymbol{x}_i),
\quad \sum\limits_{i=1}^{n_\mathcal{I}} V_i \approx \vert\mathcal{S}(\Phi_{\Lambda,\mathcal{I}})\vert.
\end{equation}
Computing the weights for each cell $K_\Omega$ is a pre-processing step that only has to be done once for each
computational mesh and well geometry.

\section{Numerical experiments and discussion}
\label{sec:numeric}

We present numerical experiments using the presented method in different setups.
All experiments are conducted with constant fluid density $\rho = \SI{1000}{\kg\per\cubic\meter}$ and viscosity $\mu = \SI{1e-3}{\Pa\second}$.
The well pressure is constant, $p_\omega = \SI{1e6}{\Pa}$, and the well radius is $r_\omega = \SI{0.1}{\meter}$ if not specified otherwise.
The permeability tensor is given as
\begin{equation}
K(\gamma_1, \gamma_2) = R_1(\gamma_1) R_2(\gamma_2) K_\alpha R^T_2(\gamma_2) R_1^T(\gamma_1),
   \quad K_\alpha = \left[\begin{array}{ccc} 1&0&0 \\ 0&1&0 \\ 0&0&\alpha \end{array}\right] \SI{1e-12}{\square\m},
\end{equation}
where
\begin{equation}
\label{eq:axisrotationmatrices}
  R_1(\gamma_1) = \left[\begin{array}{ccc} 1&0&0 \\ 0&\cos{\gamma_1}&-\sin{\gamma_1} \\ 0&\sin{\gamma_1}&\cos{\gamma_1} \end{array}\right],\quad
  R_2(\gamma_2) = \left[\begin{array}{ccc} \cos{\gamma_2}&0&\sin{\gamma_2} \\ 0&1&0 \\ -\sin{\gamma_2}&0&\cos{\gamma_2} \end{array}\right]
\end{equation}
are rotation matrices rotating vectors about $\vec{e}_1$, $\vec{e}_2$ by the rotation angle $\gamma_1$, $\gamma_2$, respectively,
and $\alpha$ is a given dimensionless K-anisotropy ratio $\alpha = \frac{K_{33}}{K_{11}} = \frac{K_{33}}{K_{22}}$.
The domain $\Omega_0 = [-100,100]\times[-100,100]\times[-50,150]~\si{\cubic\meter}$ is split in two regions,
$\Omega = [-100,100]\times[-100,100]\times[0,100]~\si{\cubic\meter}$ and $\Omega_D = \Omega_0\setminus\Omega$.
The well center-line $\Lambda$ is given by the line through the origin and $\vec{\psi} = R_1(\beta_1) R_2(\beta_2) \vec{e}_3$,
where $R_1$, $R_2$ are given in~\cref{eq:axisrotationmatrices} and $\beta_1$, $\beta_2$, are rotation angles.
The analytical solution for all cases is given in~\cref{eq:ana_w}, $q = \SI{1}{\kg\per\s\per\m}$, and $L = \vert \Lambda \cap \Omega \vert$~(in~\si{\meter}).
For all setups the inner kernel radius is chosen as $\varrho_i = f$.
In all of $\Omega_D$ and on the boundary $\partial\Omega$ the analytical solution is enforced by Dirichlet constraints, modeling the
infinite well.
The computational mesh $\Omega_h$ is a structured grid composed of regular hexahedra $K_\Omega$.
Furthermore, we define two error measures.
\begin{equation}
E_p = \frac{1}{p_\omega}\left[\frac{1}{\vert \Omega_h \vert}\sum\limits_{K_\Omega \in \Omega_h} \vert K_\Omega \vert \left( p_{e,\vec{x}_{K_\Omega}} - p_{K_\Omega} \right)^2 \right]^{\frac{1}{2}}
\end{equation}
is the relative discrete $L^2$-norm of the pressure, where $p_{e,\vec{x}_{K_\Omega}}$
is the exact pressure evaluated at the cell centroid $\vec{x}_{K_\Omega}$ and $p_{K_\Omega}$ the discrete numerical cell pressure, and
\begin{equation}
E_q = \frac{1}{q}\left[ \frac{1}{\vert \Lambda \cap \Omega_h \vert} \sum\limits_{\substack{K_\Omega \in \Omega_h \\ K_\Omega \cap \Lambda \neq \emptyset}} \vert \mathcal{I} \vert \left( q - \frac{Q_\mathcal{I}}{\vert \mathcal{I}\vert \zeta} \right)^2 \right]^{\frac{1}{2}}
\end{equation}
is the relative discrete $L^2$-norm of the source term, where $\vert \mathcal{I} \vert = \vert K_\Omega \cap \Lambda \vert$
is the length of the intersection of cell $K_\Omega$ and the well center-line $\Lambda$, $Q_\mathcal{I}$ is the discrete source term
given in~\cref{eq:discretesourceapprox}.
All setups are implemented in \dumux~\citep{flemisch2011dumu}, an open-source porous media simulator based on~\dune~\citep{dunegridpaperI:08,dunegridpaperII:08}.

\subsection{Grid convergence for different anisotropy ratios}
\label{seq:convergence}

\begin{figure}[t]
 \centering
 \includegraphics[width=1.0\textwidth]{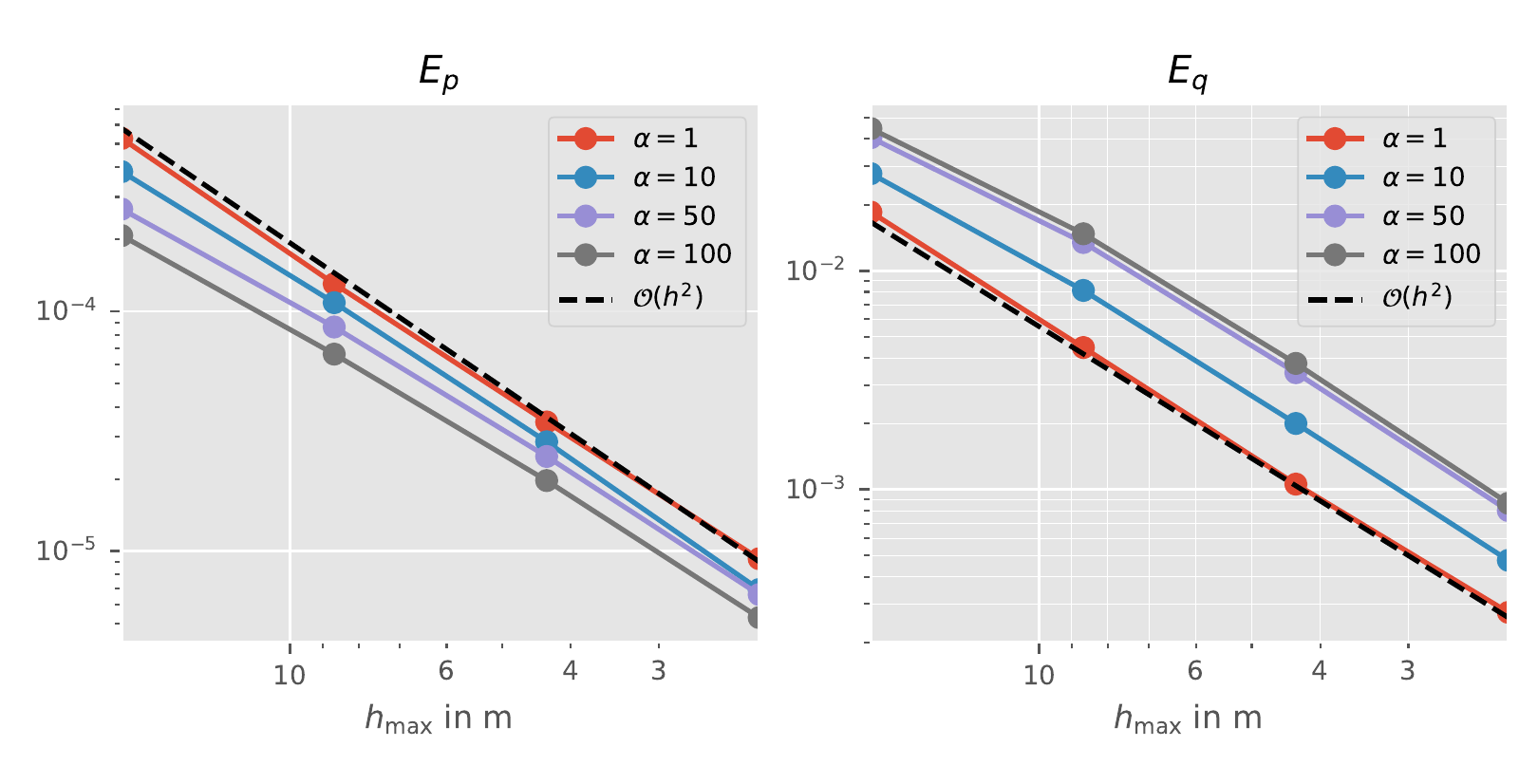}
 \caption{Grid convergence for the relative discrete $L^2$-norm pressure $E_p$ and source $E_q$ for different anisotropy ratios $\alpha$.
          None of the principal axis of the permeability tensor is aligned with the slanted well axis $\psi$ or any of the grid axes.}
 \label{fig:convergence}
\end{figure}
In the first numerical experiment grid convergence is investigated for different anisotropy ratios $\alpha$.
To this end, $h_\text{max} := \operatorname{max}\limits_{K_\Omega \in \Omega_h} h_{K_\Omega}$,
where $h_{K_\Omega}$ is defined as the maximum distance between two vertices of the cell $K_\Omega$. Starting at
a grid resolution for $\Omega_h$ of $20\times20\times10$ cells ($h_\text{max} = 10\sqrt{3}~\si{\m}$), the grid is refined uniformly.
\Cref{fig:convergence} shows the errors $E_p$ and $E_q$ for different grid resolutions and values of $\alpha$,
for $\beta_1 = \beta_2 = 20^\circ$ and $\gamma_1 = \gamma_2 = -20^\circ$, so that $K$ is a full tensor and none of the
principal axis of $K$ is aligned with the well direction. For all $\alpha$, the method shows second order
convergence for the pressure in the given norm, as expected for the MPFA\nobreakdash-O method~\citep{Schneider2018} (super convergence at cell centers).
The source term $q$ is a linear function of the pressure $p$ and also exhibits second order convergence.
We note that the errors for different $\alpha$ are not directly comparable since the analytical solution for $p$
changes with $\alpha$, although $q$ is constant.
\begin{table}
\centering
\begin{tabular}{lllll}
\toprule
\multicolumn{1}{c}{}&\multicolumn{4}{c}{$h_\text{max}$} \\
\cmidrule(r){2-5}
$\alpha$ & \SI{17.32}{\m} & \SI{8.66}{\m} & \SI{4.33}{\m} & \SI{2.17}{\m}\\
\midrule
1 & 2.0545 & 2.0724 & 1.9454 & -\\
10 & 1.7715 & 2.0184 & 2.0763 & -\\
50 & 1.5904 & 1.9747 & 2.0925 & -\\
100 & 1.5970 & 1.9666 & 2.1218 & -\\
\bottomrule
\end{tabular}
\caption{Convergence rates for $E_q$ for different anisotropy ratios $\alpha$.}
\label{tab:rates}
\end{table}
However, the convergence rates are shown to be independent of $\alpha$ with increasing grid resolution.
The convergence rates for $E_q$ (slope of the lines in~\cref{fig:convergence}) are presented in~\cref{tab:rates}.
It can be seen that rates for large grid cells and large $\alpha$ are slightly smaller. This is because the
kernel support is still under-resolved by the computational grid. For example, for $\alpha=100$, the kernel
ellipse in $x$-coordinates has major and minor axis of $a_x \approx \SI{55.9}{m}$, $b_x \approx \SI{5.6}{\m}$,
respectively, while $h_\text{max} \approx \SI{17.32}{\m}$ for the lowest grid resolution.

\subsection{Influence of the outer kernel radius $\varrho_o$}
\begin{figure}[t]
\centering
\begin{subfigure}{0.49\textwidth}
 \centering
 \includegraphics[width=1.0\textwidth]{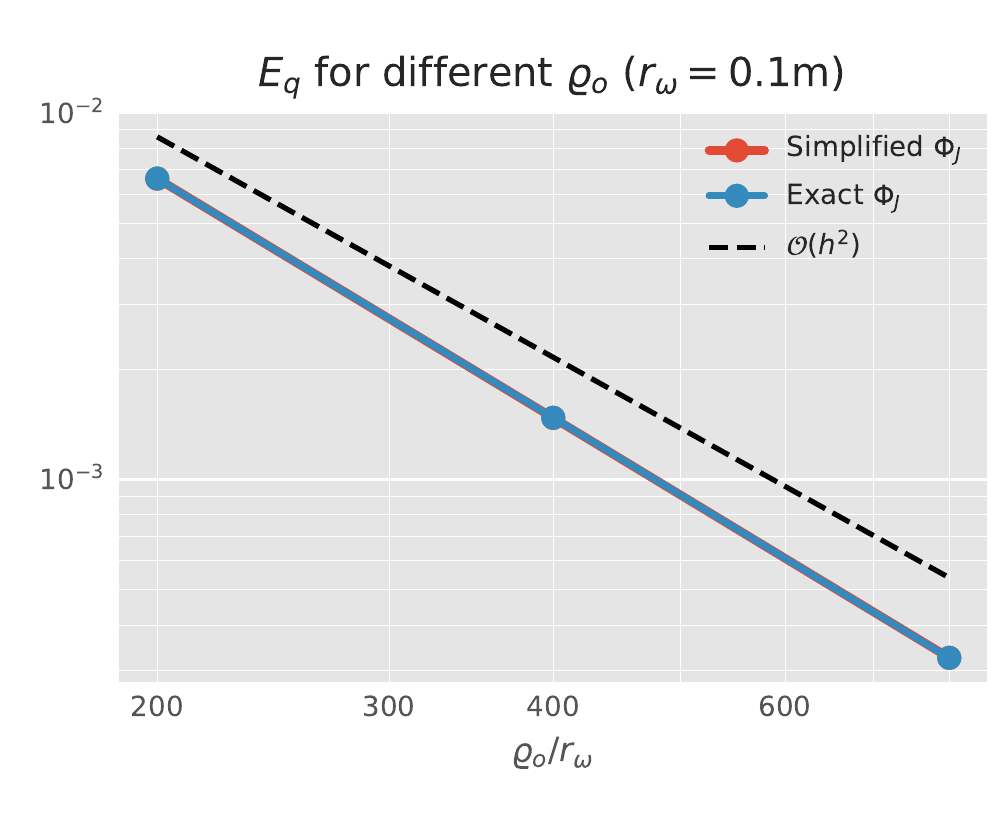}
\end{subfigure}
\begin{subfigure}{0.49\textwidth}
 \centering
 \includegraphics[width=1.0\textwidth]{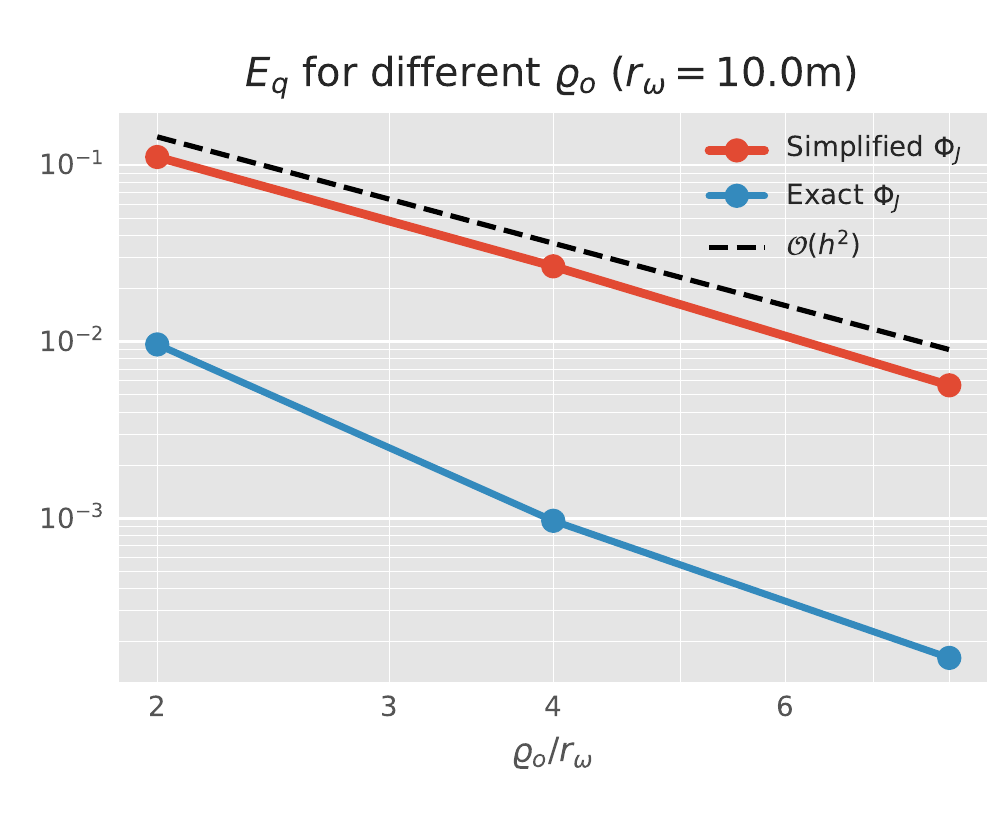}
\end{subfigure}
\caption{The source error $E_q$ for different kernel supports and the same $20\times20\times10$ computational grid ($h_\text{max} \approx \SI{17.32}{\m}$).
On the left, the case where $\varrho_o$ is only slightly larger than $r_\omega$. On the right, the case $\varrho_o \gg r_\omega$.}
\label{fig:kernelfactor}
\end{figure}
In~\citep{Koch2019a}, it is suggested that increasing the kernel support region (increasing $\varrho_o$), has
a similar effect on $E_q$ as refining the grid. However, the pressure solution is then regularized in a larger
region, so that there is a trade-off between the accuracy of the source term and the accuracy of the pressure
field with respect to the unmodified problem ($\varrho_o \rightarrow \varrho_i, \Phi_\Lambda \rightarrow \delta_\Lambda$).
However, every discrete cell $K_\Omega$ can be also interpreted as a kernel support region, such that
the choice of $\Phi_\Lambda$ enables us to better control the discretization error as soon as $\mathcal{S}(\Phi_\Lambda, \mathcal{I})$
becomes larger than $K_\Omega$.

As shown in~\citep{Peaceman1983} and~\cref{fig:detplot}, isobars become circular, in the transformed domain $U(\Omega)$,
with increasing distance to the well. Therefore, a reasonable simplification is $\Phi_J \approx 4$ if $\varrho_o \gg r_\omega$.
This is completely analogous to the assumption of circular isobars in~\citep{Peaceman1983}, where an estimate
of the error introduced by the assumptions is given for the two-dimensional case.

In the following numerical experiment, we step-wise increase the kernel radius $\varrho_o$, for the same
$20\times20\times10$ grid. This is done once for the case, where $\varrho_o \gg r_\omega$ and for the case for which
$\varrho_o$ is only slightly larger than $r_\omega$.
Furthermore, $\beta_1 = \beta_2 = 20^\circ$ and $\gamma_1 = \gamma_2 = -20^\circ$. The results are
shown in~\cref{fig:kernelfactor}. First, it can be seen that doubling $\varrho_o$ leads to a $4$-times smaller
error $E_q$. This can be explained by the fact that the larger the kernel, the more grid cells resolve the
kernel support, and the better is the approximation of $p_0$.
Furthermore, the result is consistent with the results in~\citep{Koch2019a}.
Moreover, \cref{fig:kernelfactor} suggests that for $\varrho_o \gg r_\omega$ the simplification of the kernel function
($\Phi_J \approx 4$) is not visible in $E_q$, while for kernel radii slightly smaller than the well radius,
the simplification increases $E_q$ by an order of magnitude in comparison to the case using the exact kernel
function as derived in~\cref{sec:anisokernel}.
The results show that the presented method is also applicable in cases where the grid resolution is very
close to the well radius. An adaption of the presented method for other applications, such as the simulation
of flow in vascularized tissue, where such ratios of vessel radius to cell size are typical, cf.~\citep{Koch2019a},
is therefore well-conceivable.

\subsection{Robustness with respect to rotation}
\begin{figure}[tb]
\centering
\begin{subfigure}{0.49\textwidth}
 \centering
 \includegraphics[width=1.0\textwidth]{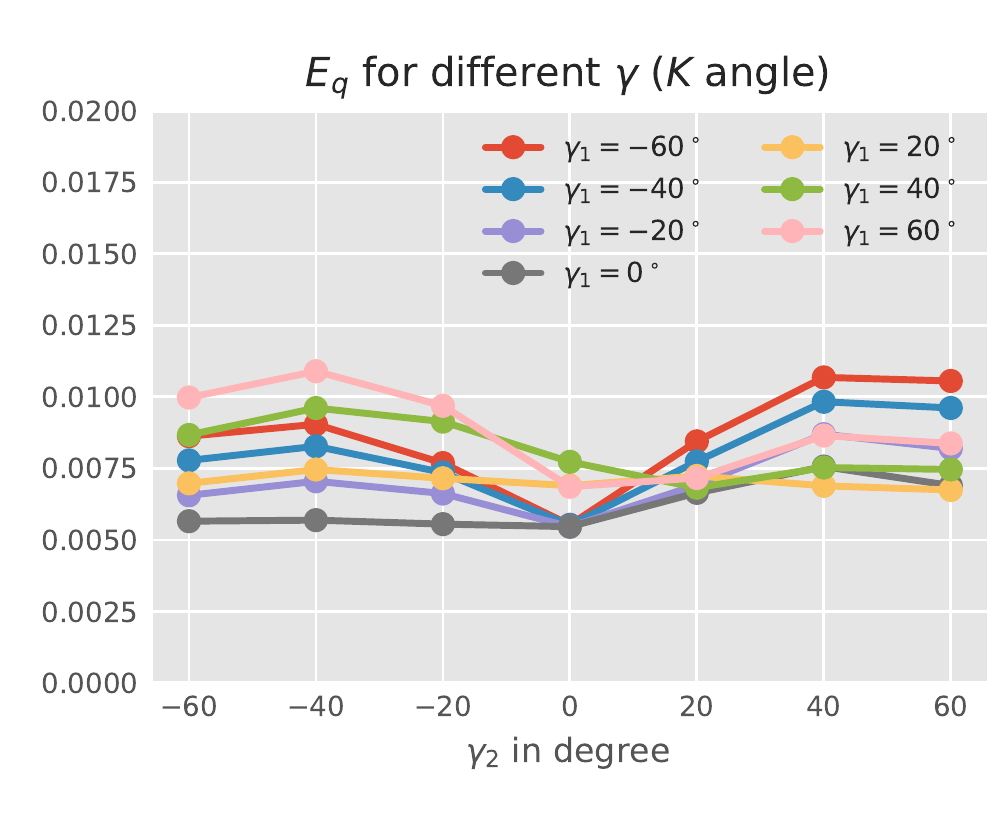}
\end{subfigure}
\begin{subfigure}{0.49\textwidth}
 \centering
 \includegraphics[width=1.0\textwidth]{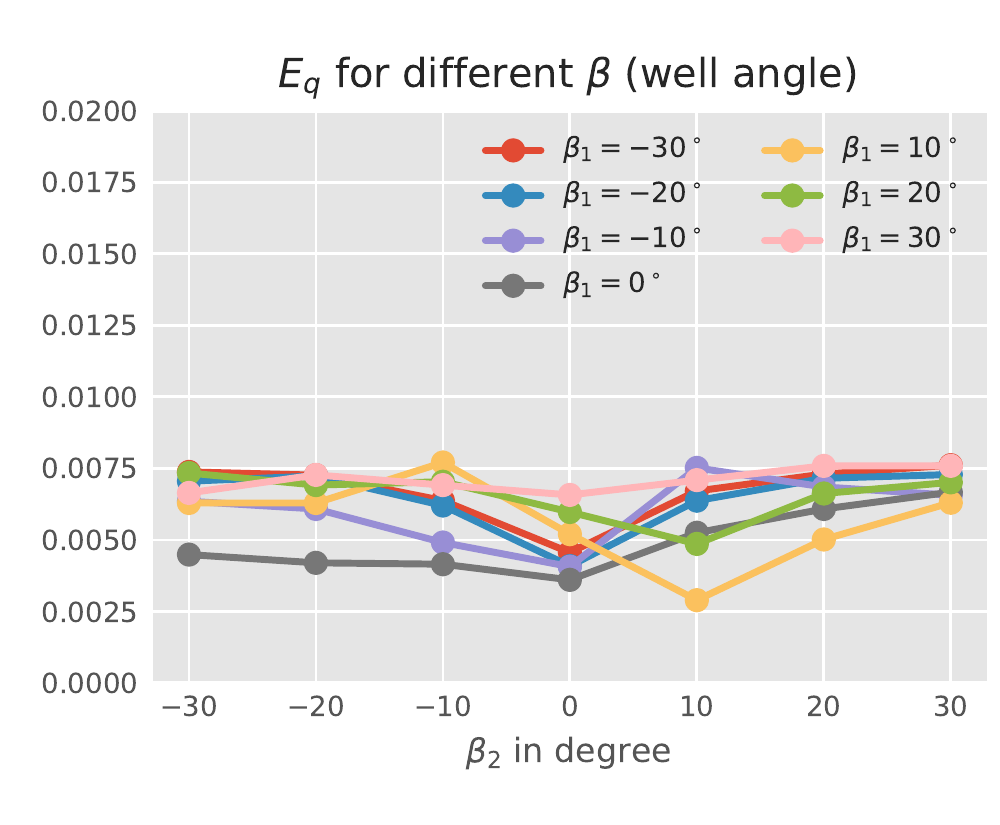}
\end{subfigure}
\caption{The source error $E_q$ for rotations of the permeability tensor (left) and different well orientations (right).}
\label{fig:angles}
\end{figure}

In the following numerical experiment, we use a single computational mesh with a given resolution for $\Omega_h$: $20\times20\times10$.
First, the well direction is fixed, and the permeability tensor is rotated by varying $\gamma_1$ and $\gamma_2$. Then the permeability
tensor is fixed and the well is rotated by varying $\beta_1$ and $\beta_2$. The results are shown in~\cref{fig:angles}.
It can be seen that the presented well model is rather robust with respect to rotations. Possible effects influencing the
approximation error $E_q$, include the different quality of the kernel integral for different angles with respect
to the grid axes, and differences in the flux approximation quality
of the MPFA\nobreakdash-O method depending on the face co-normal $\vec{d}_{K_\Omega,\sigma} = K\vec{n}_{K_\Omega,\sigma}$.
Additionally, for different well angles the number and size of intersections
$K_\Omega \cap \Lambda$ can have an influence on the discrete error.

\subsection{Comparison with a Peaceman-type well model}
\label{sec:peaceman}

In particular for petroleum engineering applications, commercial codes typically use Peaceman-type well models~\citep{eclipse2014,imex2014}.
In~\citep{Peaceman1983}, Peaceman extended his well known well-index-based well model
for anisotropic diagonal permeability tensors and non-cubic
but structured rectangular grids. The discrete source term in a computational cell $K_\Omega$ is approximated by
\begin{align}
\label{eq:peaceman}
Q_{K_\Omega} &= 2\pi\frac{\rho}{\mu}(p_\omega - p_0)L_{K_\Omega} \frac{\sqrt{K_{11} K_{22}}}{\ln{\left(\frac{r_0}{r_\omega}\right)}},\\
r_0 &= \frac{e^{-\gamma}}{2}\frac{\left[ \left( \frac{K_{22}}{K_{11}} \right)^{\frac{1}{2}}\Delta x^2 + \left( \frac{K_{11}}{K_{22}} \right)^{\frac{1}{2}} \Delta y^2 \right]^{\frac{1}{2}}}{ \left( \frac{K_{11}}{K_{22}} \right)^{\frac{1}{4}} + \left( \frac{K_{22}}{K_{11}} \right)^{\frac{1}{4}} },
\end{align}
where $\Delta x$ and $\Delta y$ are the horizontal dimensions of the cell containing the well,
$L_{K_\Omega} = \vert\mathcal{I}\vert$ the length of the well segment contained in $K_\Omega$, and $\gamma$ the Euler--Mascheroni constant.
The Peaceman model has several known limitations. Its derivation only applies to K-orthogonal structured grids, where the well
is oriented along one of the grid axes, and perfectly horizontally centered within a vertical column of computational cells $K_\Omega$.
Furthermore, the derivation is specific to cell-centered finite difference schemes with 5-point stencil.
Moreover, computational cells may have to be significantly
larger than the well radius (depending on the degree of anisotropy) for optimal accuracy.
The Peaceman model has been generalized for slanted wells with arbitrary orientation, for example in~\citep{Alvestad1994}. The
Alvestad model~\citep{Alvestad1994} has been adapted for finite volumes, for example in~\citep{Aavatsmark2003}
(formula given in~\cref{sec:appendixE}, subsequently referred to as \textsc{pm} well model).
Such extensions usually constitute a
reasonable directional weighting of the original Peaceman model but are
not directly derived from the mathematical analysis of the underlying problem~\citep{Aavatsmark2003}.

The herein presented model has none of the above-mentioned limitations. In particular, the presented model is valid for arbitrary
positive definite and symmetric permeability tensors, unstructured grids, and is independent of the discretization scheme. Moreover,
the presented model is consistent and we show grid convergence in the numerical experiments in~\cref{seq:convergence}.
However admittedly, the Peaceman-type models are cell-local, thus computationally cheaper and easier to implement.

Several limitations of the Peaceman well models make it difficult to fairly compare it with our new model.
For cases for which all assumptions of Peaceman are valid, our numerical studies (not shown here) suggest that the Peaceman well model
is generally superior to the presented model with distributed sources. This is because it takes the analytical solution as well as the
spatial discretization method into account. For cases where some assumptions are violated, for example off-center wells or slanted wells,
it is difficult to construct cases where the analytical solution is readily constructed but does not feature a singularity on the boundary.
Our preliminary numerical studies for such cases
(for example the slanted well case in~\cref{seq:convergence} without rotation of the permeability tensor) show large deviations
($>\SI{10}{\percent}$ error in total source term) from the analytical solution for the \textsc{pm} well model.
However, these errors may be distorted by errors made in the discrete approximation of the singular boundary condition,
where the well intersects the boundary.
Finally, for the general case of unstructured grids, simplex grids, and full permeability tenors it is unclear how to apply the original Peaceman model.
However, we know that the presented method is consistent (at least for a single straight well), and thus, the numerical solution converges to the
exact solution with grid refinement. Therefore, we expect that the numerical solution on a very fine grid using the distributed source
model is a reasonable reference solution.

\begin{figure}[t]
 \centering
 \includegraphics[width=1.0\textwidth]{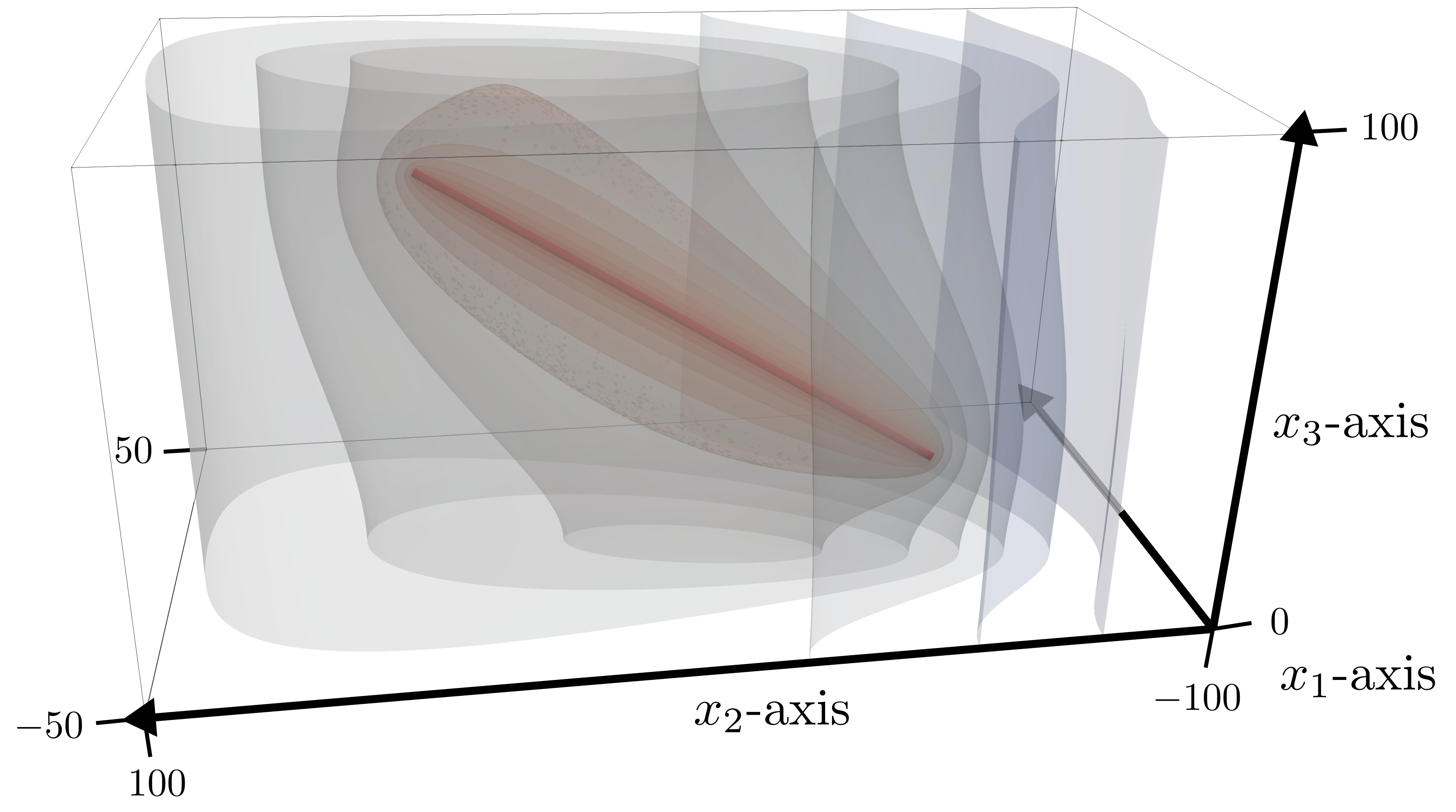}
 \caption{The computational domain for the comparison with a Peaceman-type well model. The well is visualized with a $10$-fold increased radius.
          A selection of pressure iso-surfaces of the reference solution are shown with reduced opacity. The domain extent is given in units of \si{\m}.}
 \label{fig:comparison_domain}
\end{figure}

\begin{figure}[t]
 \centering
 \includegraphics[width=0.9\textwidth]{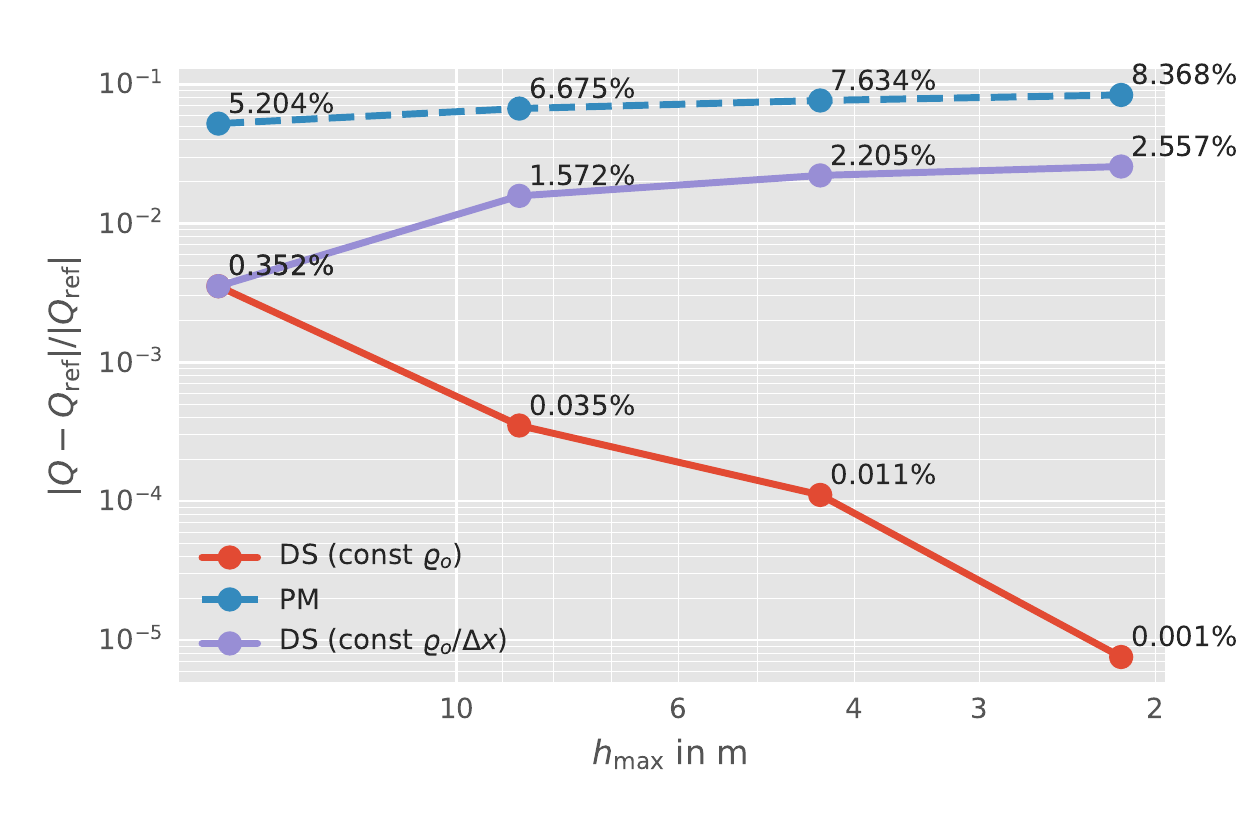}
 \caption{Comparison of the relative integral source error between a Peaceman-type model (\textsc{pm}) and the new model (\textsc{ds}) for various grid refinements.
          The error is computed with respect to a reference solution $Q_\text{ref}$. Both axes are logarithmic.}
 \label{fig:comparison}
\end{figure}

We compare our model to the \textsc{pm} well model in a numerical experiment.
The computational domain $\Omega = [-50,50]\times[-100,100]\times[0,100]~\si{\cubic\meter}$
contains a slanted straight well $\Lambda$
with end points at $\vec{x}_{\Lambda,1} = [-20, -50, 25]^T~\si{\meter}$, $\vec{x}_{\Lambda,2} = [20, 50, 75]^T~\si{\meter}$.
The permeability tensor is a diagonal tensor $K(\gamma_1,\gamma_2)$, with $\gamma_1 = 0^\circ, \gamma_2 = 90^\circ, \alpha = 0.1$.
The structured cube grid $\Omega_h$ is successively, uniformly refined starting with $10\times20\times10$ cells ($h_\text{max} \approx \SI{17.32}{\m}$).
The well radius is $r_\omega = \SI{0.1}{\m}$ ($\Delta x / r_\omega = 200$ for the coarsest grid).
The kernel support region (chosen as $\varrho_o / r_\omega = 100$) only extends over few cells in the coarsest grid,
so that the regularization effect is minimized. On the boundary $\partial \Omega$, we
specify Neumann no-flow boundary conditions, that is $(K \grad p) \cdot \vec{n} = 0$, except for the planes perpendicular to the $x_1$-axis, where the Dirichlet boundary condition $p_D(x_2 = -100) = \SI{1e5}{\Pa}$, $p_D(x_2 = 100) = \SI{3e5}{\Pa}$ are enforced.
The reference solution is computed
with $160\times320\times160$ cells ($h_\text{max} \approx \SI{1.08}{\m}$).
The computational domain with pressure iso-surfaces of the reference solution are shown in~\cref{fig:comparison_domain}.
In~\cref{fig:comparison}, the relative integral source error
\begin{equation}
E_Q = \frac{\vert Q - Q_\text{ref}\vert}{\vert Q_\text{ref} \vert}, \quad Q = \sum\limits_{K_\Omega \in \Omega_h} Q_{K_\Omega},
\end{equation}
with respect to the reference
solution $Q_\text{ref}$ is shown for grids with different refinement. In a variant of the distributed source model (\textsc{ds}), the extent of the
kernel support is adapted to the grid size. This is to keep the regularization effect of the kernel function minimal in order to get, in
addition to a good approximation of the source term, a better approximation of the pressure solution close to the well.
For $\varrho_o / r_\omega = 100$, the extent of the
kernel ellipse $E_{\Phi,x}$  is given by its major and minor axes,
$\SI{16.12}{\m}$ and $\SI{12.54}{\m}$. For the reference solution this extent is kept constant with grid refinement. While this ensures
a very good approximation of the source term, the pressure solution is regularized in a larger neighborhood of the well. In the variant,
the kernel support is adapted proportional to $h_\text{max}$,
so that for the finest grid shown in~\cref{fig:comparison} ($80\times160\times80$ cells, $h_\text{max} \approx \SI{2.17}{\m}$),
the $E_{\Phi,x}$ major and minor axes measure $\SI{2.01}{\m}$ and $\SI{1.57}{\m}$.
It is evident that the numerical solution for the distributed source model converges to the reference solution. More importantly,
the relative error is small ($<\SI{0.5}{\percent}$) even for the coarsest grid. In comparison, the difference to the Peaceman-type model is
large ($>\SI{5}{\percent}$). In particular, the error grows with grid refinement (to $>\SI{8}{\percent}$), signifying that the generalization of
Peaceman's model for arbitrarily-oriented wells is not consistent. The result is comparable with the observations in~\citep[][Table 2]{Aavatsmark2003},
where Alvestad's well indices are compared to a new numerically computed well index, and it is shown that the difference between those
two well indices grow, the higher the $r_\omega / \Delta x$ ratio. In the variant of the \textsc{ds} model, the error in the source term
with respect to the reference solution also grows with larger $r_\omega / \Delta x$ ratio. However, the error is consistently smaller (by a factor $>3$)
than for the \textsc{pm} well model.
\begin{figure}[t]
 \centering
 \includegraphics[width=1.0\textwidth]{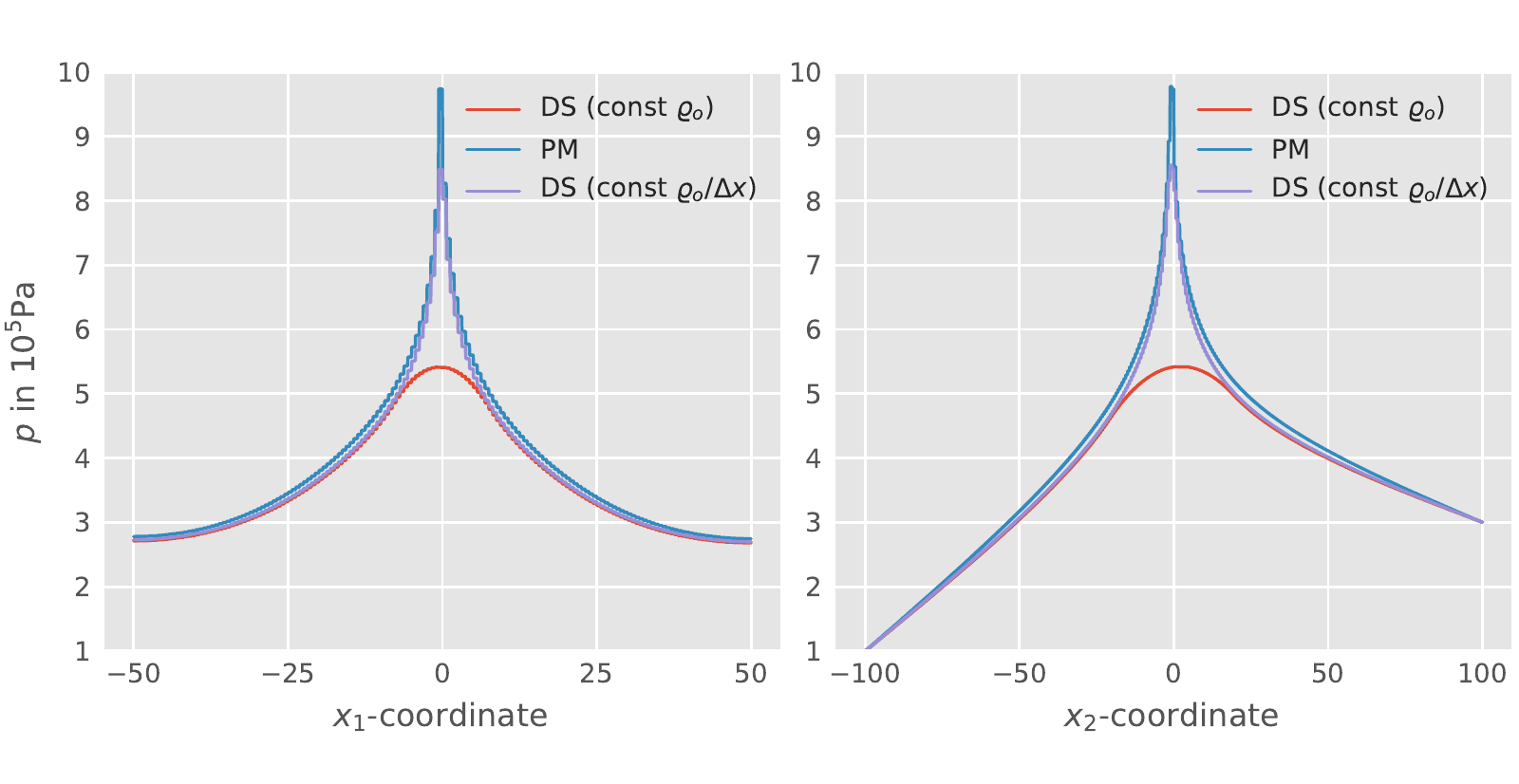}
 \caption{Numerical pressure solutions plotted along the $x_1$ and the $x_2$-axis for a Peaceman-type model (\textsc{pm}) and the new model (\textsc{ds}) for a grid resolution
          of $160\times320\times160$ cells.}
 \label{fig:comparison_pressure}
\end{figure}
\Cref{fig:comparison_pressure} shows the numerical pressure solutions along the $x_1$ and the $x_2$-axis, for
the reference grid resolution. It can be clearly seen that for the reference solution (\textsc{ds}) the pressure solution is regularized. For
the variant of \textsc{ds}, the regularization is minimized, however in the far field the solution matches the reference solution better than
the \textsc{pm} well model, which is due to the better approximation of the source term (see~\cref{fig:comparison}).
We also note that the regularized solution leads to an altered solution in the near-field of the well but to a better approximation of
the source term and thus the far field pressure (outside the kernel support),
whereas the poor approximation of the source term in the \textsc{pm} method leads to a globally poor pressure solution.

\section{Summary}

A new well model was presented for which the mass exchange between a well and an embedding porous medium is modeled with
a source term spatially distributed by a local kernel function. In the spirit of well-index-based well models the source term for a well
with given bottom hole pressure is computed based on the numerical pressure in cells intersecting the well. However, the presented derivation of the new model
is independent of the discretization method and the type of computational grid. The new model was shown to be consistent in a numerical experiment
and exhibited grid convergence with the expected rates. In the same experiment it is shown that the absolute error with respect
to an analytical solution is relatively small, even for coarse computational grids and small kernel support.
It was shown, that the error in the source term can be decreased by increasing
the region over which the source term is distributed. However, coincidentally, the pressure profile close to the well (inside the kernel support)
becomes increasingly regularized. A comparison with a Peaceman-type well model generalized for arbitrarily oriented wells,
suggested that even if the region is chosen to be very small (only covering the neighboring cells of cells with well intersection), thus minimizing the regularization
effect, the source term can be approximated with good accuracy ($<$~\SI{2}{\percent} error with respect to a reference solution),
whereas the Peaceman-type well model for the same case showed larger differences ($>$~\SI{8}{\percent})
which also had a negative global effect on the pressure solution. The example showed that the new model gives the choice between a more accurate representation
of the near-well pressure and a more accurate representation of the source term. Additionally, it was shown that if the source term is accurately approximated,
the far-field pressure (outside kernel support) is equally well-approximated while the regularization of the pressure profile only happens locally in the well neighborhood.
On the other hand, a bad approximation of the source term leads to global errors in the pressure profile.
Finally, the new model was shown to be robust with respect to well rotation, as well as robust with respect to rotations of the anisotropic permeability tensor.

In this work, the well model derivation is restricted to one-phase flow and possible modifications for multi-phase flow are yet to be explored.
An extension of the well model for wells with casing is easily conceivable, combining the findings in this work with the derivations presented in~\citep{Koch2019a}.
Moreover, the herein presented cases considered the case of a given constant bottom hole pressure. However, the results of~\citep{Koch2019a} indicate,
that the presented model may be extended for the coupled 1d-3d case where the well pressure solves an additional one-dimensional
partial differential equation in the well domain. Finally, using the superposition principle as shown in~\citep{Koch2019a}, which equally applies for
the case of anisotropic permeabilities, the presented well model is also expected to provide good approximations when multiple wells are present in the domain.

\section*{Acknowledgements}

This work was financially supported by the German Research Foundation (DGF), within the Cluster of Excellence in Simulation Technology (EXC 310),
and the Collaborative Research Center on Interface-Driven Multi-Field Processes in Porous Media (SFB 1313, Project Number 327154368).

\begin{appendices}
\crefalias{section}{appsec}

\section{Rodrigues' rotation formula}
\label{sec:appendixA}
We want to rotate a given basis $\mathcal{B} = \{ \vec{e}_1, \vec{e}_2, \vec{e}_3 \}$ such that $\vec{e}_3$ is aligned with a vector $\vec{\psi}$.
The Rodrigues' rotation formula~\citep{dai2015} describes a rotated vector $\vec{x}_\text{rot}$ obtained by rotating a vector $\vec{x}$ by the angle $\theta$ about an axis given by the unit normal vector $\vec{k}$
\begin{equation}
\vec{x}_\text{rot} = \vec{x} \cos{\theta} + (\vec{k} \times \vec{x}) \sin{\theta} + \vec{k}(\vec{k}\cdot\vec{x})(1 - \cos{\theta}).
\end{equation}
The desired rotation can be described by a rotation by an angle $\theta = \pi$ about the axis $\vec{k} = \frac{(\vec{e}_3 + \vec{\psi})}{2\vert\vec{e}_3 + \vec{\psi}\vert}$,
\begin{equation}
\label{eq:rodriguespi}
\vec{x}_{\text{rot},\pi} = 2\vec{k}(\vec{k}\cdot\vec{x}) -\vec{x}.
\end{equation}
\Cref{eq:rodriguespi} can be expressed in matrix notation as $\vec{x}_{\text{rot},\pi} = R\vec{x}$ with
\begin{equation}
R = \left(2 \vec{k}\vec{k}^T - I\right), \quad R = R^T = R^{-1}, \quad \operatorname{det}(R) = 1.
\end{equation}

\section{Transformation of Laplace operator}
\label{sec:appendixB}
The Joukowsky transformation is a complex function $z = T^{-1}(w)$ that can be decomposed in
its real and imaginary parts, $x = \Re{(z)}$ and $y = \Im{(z)}$. Furthermore, let $u = \Re{(w)}$ and $v = \Im{(w)}$.
As a conformal mapping, $z$ satisfies the Cauchy Riemann equations~\citep{Nehari1975}
\begin{equation}
\label{eq:cauchyriemann}
\frac{\partial x}{\partial u} = \frac{\partial y}{\partial v} \quad \text{and} \quad \frac{\partial x}{\partial v} = -\frac{\partial y}{\partial u}.
\end{equation}
The Jacobian of the transformation $J_{T^{-1}}$ is given by~\cref{eq:jou_jac}.
We investigation the effect of the transformation on the Laplace operator
\begin{equation}
  \Delta_w p = \frac{\partial^2 p}{\partial u^2} + \frac{\partial^2 p}{\partial v^2},
\end{equation}
where $p$ is analytic in $\Omega_w$.
Applying the chain rule yields
\begin{align}
  \frac{\partial p}{\partial u} &= \frac{\partial p}{\partial x}\frac{\partial x}{\partial u} + \frac{\partial p}{\partial y}\frac{\partial y}{\partial u}, \\
  \frac{\partial^2 p}{\partial u^2} &= \frac{\partial p}{\partial x}\frac{\partial^2 x}{\partial u^2} + \frac{\partial p}{\partial y}\frac{\partial^2 y}{\partial u^2}
                                     + \frac{\partial}{\partial x}\left(\frac{\partial p}{\partial u}\right)\frac{\partial x}{\partial u}
                                     + \frac{\partial}{\partial y}\left(\frac{\partial p}{\partial u}\right)\frac{\partial y}{\partial u} \\\nonumber
                                    &= \frac{\partial p}{\partial x}\frac{\partial^2 x}{\partial u^2} + \frac{\partial p}{\partial y}\frac{\partial^2 y}{\partial u^2}
                                     + \frac{\partial^2 p}{\partial x^2}\left( \frac{\partial x}{\partial u} \right)^2
                                     + 2\frac{\partial^2 p}{\partial x \partial y}\frac{\partial x}{\partial u}\frac{\partial y}{\partial u}
                                     + \frac{\partial^2 p}{\partial y^2}\left( \frac{\partial y}{\partial u} \right)^2.
\end{align}
Analogously, we arrive at a similar expression for $\partial^2 p / \partial v^2$. Using~\cref{eq:cauchyriemann} and
\begin{equation}
\frac{\partial^2 y}{\partial v^2} = \frac{\partial}{\partial v}\left(\frac{\partial y}{\partial v}\right)
                                  \overequal{(B.1)} \frac{\partial}{\partial v}\left(\frac{\partial x}{\partial u}\right)
                                  = \frac{\partial}{\partial u}\left(\frac{\partial x}{\partial v}\right)
                                  \overequal{(B.1)} -\frac{\partial y^2}{\partial u^2}, \quad \frac{\partial^2 x}{\partial v^2} = -\frac{\partial^2 x}{\partial u^2},
\end{equation}
we find that
\begin{equation}
\frac{\partial^2 p}{\partial u^2} + \frac{\partial^2 p}{\partial v^2}
  = \left[ \left( \frac{\partial x}{\partial u} \right)^2 + \left( \frac{\partial y}{\partial u} \right)^2 \right] \left[ \frac{\partial^2 p}{\partial x^2} + \frac{\partial^2 p}{\partial y^2} \right].
\end{equation}
With the complex derivative of $T^{-1}$~\citep{Rudin1987},
\begin{equation}
\frac{\partial T^{-1}}{\partial w} = \frac{\partial T^{-1}}{\partial u} = \frac{\partial x}{\partial u} + i\frac{\partial y}{\partial u}  \quad \text{and} \quad \left| \frac{\partial T^{-1}}{\partial w} \right| = \sqrt{ \left( \frac{\partial x}{\partial u} \right)^2 + \left( \frac{\partial y}{\partial u} \right)^2}.
\end{equation}
From the determinant of the Jacobian of the transformation, we find
\begin{equation}
\operatorname{det}(J_{T^{-1}}) = \frac{\partial x}{\partial u} \frac{\partial y}{\partial v} - \frac{\partial x}{\partial u}\frac{\partial y}{\partial v} = \left(\frac{\partial x}{\partial u} \right)^2 + \left( \frac{\partial y}{\partial u} \right)^2 = \left| \frac{\partial T^{-1}}{\partial w}\right|^2,
\end{equation}
using~\cref{eq:cauchyriemann}. Hence,
\begin{equation}
\label{eq:laplace_conformal_map}
\Delta_w p = \left| \frac{\partial T^{-1}}{\partial w}\right|^2 \Delta_z p = \left|\operatorname{det}(J_{T^{-1}})\right| \Delta_z p,
\end{equation}
which also proofs that any harmonic function ($\Delta f = 0$) yields another harmonic function after a coordinate transformation with a conformal mapping. \qed

\section{Source scaling factor in $w$-coordinates}
\label{sec:appendixC}
We want to construct a pressure solution in $w$-coordinates such that the total mass flux over the well boundary
matches the specified boundary condition in $x$-coordinates.
Hence, the total mass flux over the boundary of a well segment with length $\hat{L}$ in $w$-coordinates needs to match the total mass
flux over the boundary of a well segment with length $L$ in $x$-coordinates.
A $\hat{q}$ has to be chosen such that $qL = \hat{q}\hat{L}$. A relation between $L$ and $\hat{L}$ can be
derived by looking at two related volume integrals. The Joukowsky transformation only affects the two-dimensional well-bore
plane such that length $\hat{L}$ of a well segment is not affected.
The volume of that well segment in $v$-coordinates (an elliptic cylinder) is given by
\begin{equation}
V_v = \pi ab \hat{L}.
\end{equation}
The volume of the same well segment in $x$-coordinates (a circular cylinder with slanted parallel planar elliptic caps)
is given by
\begin{equation}
V_x = \vert \vec{\psi}^T\tilde{S}\tilde{R}^T\vec{e}_3 \vert \vert E_{\omega,x} \vert L = \pi r_\omega^2 L,
\end{equation}
where
$\vert E_{\omega,x} \vert$ is the area of the well-bore ellipse described by~\cref{eq:wellboreellipse} transformed to $x$-coordinates
(as shown in~\cref{fig:transformation}) and the last
equality uses the fact that the integral can be transformed to an integral over a regular cylinder with radius $r_\omega$ and length $L$.
From the transformation theorem, we know that $V_x = V_v \operatorname{det}(\tilde{S})$. As the parameter $k_I$
is chosen such that $\operatorname{det}(\tilde{S}) = 1$,
\begin{equation}
\hat{L} = L\frac{r_\omega^2}{ab},
\end{equation}
and if the source term is chosen as
\begin{equation}
\hat{q} = q\frac{ab}{r_\omega^2} := q\zeta,
\end{equation}
then $qL = \hat{q}\hat{L}$. \qed

\section{Determinant of the Joukowsky transformation}
\label{sec:appendixD}
In~\cref{sec:appendixB}, we show that $\left|\operatorname{det}(J_{T^{-1}})\right| = \left| \frac{\partial z}{\partial w}\right|^2$.
Using complex differentiation,
\begin{align}
\frac{\partial z}{\partial w} = \frac{\partial z}{\partial u} = \frac{\partial}{\partial u}\left[ \frac{1}{2}\left( w + \frac{f^2}{w} \right)\right]
=  \frac{1}{2}\left( 1 - \frac{f^2}{w^2} \right) = \frac{1}{2}\left( 1 - \frac{f^2 \overline{w}^2}{\vert w\vert^4} \right),
\end{align}
where we used the identities $w^{-1} = \overline{w}\vert w\vert^{-2}$, $\overline{w}$ denoting the complex conjugate of $w$, and $\overline{w^2} = \overline{w}^2$.
Furthermore,
\begin{align}
\Re{\left(\frac{\partial z}{\partial w}\right)}^2 &= \frac{1}{4}\left( 1 - \frac{2f^2 \Re{\left(w^2\right)}}{\vert w\vert^4} + \frac{f^4\Re{\left(w^2\right)}^2}{\vert w\vert^8}\right), \\
\Im{\left(\frac{\partial z}{\partial w}\right)}^2 &= \frac{1}{4}\left(\frac{f^4 \Im{\left(w^2\right)}^2}{\vert w\vert^8}\right), \quad \Re{\left(w^2\right)}^2 + \Im{\left(w^2\right)}^2 =  \vert w\vert^4,
\end{align}
such that
\begin{equation}
\left| \frac{\partial z}{\partial w}\right|^2 = \Re{\left(\frac{\partial z}{\partial w}\right)}^2 + \Im{\left(\frac{\partial z}{\partial w}\right)}^2
= \frac{1}{4}\left( 1 + \frac{f^4 - 2f^2\Re(w^2)}{\vert w \vert^4} \right).
\end{equation}
\qed

\section{Extension of Peaceman well model for slanted wells}
\label{sec:appendixE}
For a given well direction $\vec{\psi} = [\psi_1, \psi_2, \psi_3]^T$ and a cell $K_\Omega$
with dimensions $\Delta x \times \Delta y \times \Delta z$,
to obtain the generalized well model due to~\citep{Alvestad1994},
reformulated for cell-centered finite volume schemes in~\citep{Aavatsmark2003},
replace $k = \sqrt{K_{11} K_{22}}$ in~\cref{eq:peaceman} by
\begin{equation}
k = (\psi_1^2 K_{22} K_{33} + \psi_2^2 K_{11} K_{33} + \psi_3^2 K_{11} K_{22})^{\frac{1}{2}}
\end{equation}
and the expression for $r_0$ by
\begin{align}
r_0 &= \frac{e^{-\gamma}}{2}\frac{\left( \Delta L_1^2  + \Delta L_2^2 \right)^{\frac{1}{2}}}{\left(\sqrt{A_1} + \sqrt{A_2}\right)}, \quad \text{with}\\
\Delta L_1^2 &= \left( \frac{K_{22}}{K_{33}} \right)^{\frac{1}{2}} \Delta z^2 \psi_1^2
               + \left( \frac{K_{33}}{K_{11}} \right)^{\frac{1}{2}} \Delta x^2 \psi_2^2
               + \left( \frac{K_{11}}{K_{22}} \right)^{\frac{1}{2}} \Delta y^2 \psi_3^2, \\
\Delta L_2^2 &= \left( \frac{K_{33}}{K_{22}} \right)^{\frac{1}{2}} \Delta y^2 \psi_1^2
               + \left( \frac{K_{11}}{K_{33}} \right)^{\frac{1}{2}} \Delta z^2 \psi_2^2
               + \left( \frac{K_{22}}{K_{11}} \right)^{\frac{1}{2}} \Delta x^2 \psi_3^2, \\
A_1 &= \left( \frac{K_{22}}{K_{33}} \right)^{\frac{1}{2}}  \psi_1^2
        + \left( \frac{K_{33}}{K_{11}} \right)^{\frac{1}{2}} \psi_2^2
        + \left( \frac{K_{11}}{K_{22}} \right)^{\frac{1}{2}}  \psi_3^2, \\
A_2 &= \left( \frac{K_{33}}{K_{22}} \right)^{\frac{1}{2}} \psi_1^2
        + \left( \frac{K_{11}}{K_{33}} \right)^{\frac{1}{2}} \psi_2^2
        + \left( \frac{K_{22}}{K_{11}} \right)^{\frac{1}{2}} \psi_3^2.
\end{align}
We note that the formula reduces to~\cref{eq:peaceman},
if $\vec{\psi}$ is aligned with a coordinate axis.

\end{appendices}

\bibliography{wellmodel}
\bibliographystyle{elsarticle-num}

\end{document}